\newif\ifusenix
\newif\ifacm
\newif\ifmcom
\newif\ifacrosslength
\newif\ifsurface
\newif\ifshowsumm

\showsummfalse
\usenixtrue
\acmfalse
\mcomfalse
\surfacetrue

\newcommand{\paperTitle}{WiForce\xspace}
\newcommand{\name}{WiForce\xspace}
\newcommand{\notreview}[1]{{\color{red}\textbf{Do Not Review: #1}}}
\newcommand{\notdone}[1]{{\color{red}\textbf{#1}}}
\newcommand{\done}[1]{{\color{green}\textbf{#1}}}
\mcomfalse

\ifusenix
	\documentclass[letterpaper,twocolumn,10pt]{article}
	\usepackage{usenix}
	\usepackage{times}
\fi
\ifacm
  \documentclass[10pt,sigconf]{acmart}
  \renewcommand\footnotetextcopyrightpermission[1]{} 
  \setcopyright{none}
\fi

\ifmcom
  \documentclass{sig-alternate-10pt}
\fi

\usepackage{cite}
\usepackage{tikz}
\usetikzlibrary{positioning}
\usetikzlibrary{arrows}
\usetikzlibrary{shapes,snakes}
\usetikzlibrary{plotmarks}
\usepackage{pgfplots}
\usepackage{xparse}
\usepackage{xcolor}
\usepackage{amsfonts}
\usepackage{balance}
\usepackage{xspace}
\PassOptionsToPackage{hyphens}{url}\usepackage{hyperref}
\usepackage{color}
\usepackage{graphics}
\usepackage{graphicx}
\usepackage{url}
\usepackage{listings}
\usepackage{multicol}
\usepackage{multirow}
\usepackage[scaled]{helvet}
\usepackage{rotating}
\usepackage{xspace}
\usepackage{algorithm}
\usepackage{comment}
\usepackage{enumitem}
\usepackage{amsmath}
\usepackage{mathrsfs}
\usepackage{cancel}
\usepackage{cleveref}
\usepackage{wrapfig}
\usepackage[skip=0pt]{subcaption}
\usepackage{commath}
\usepackage{booktabs}
\usepackage{array}
\usepackage[bottom]{footmisc}
\usepackage[font=small,labelfont=bf]{caption}
\usepackage{import}
\pgfplotsset{compat=1.3}
\pgfplotsset{every axis/.append style={
                    ylabel style={font=\small \color{white!15!black}},
                    xlabel style={font=\small \color{white!15!black}},
                    tick label style={font=\scriptsize},
                    },
        /pgfplots/xlabel shift={-5pt}
        }

\usepackage{titlesec}

\titlespacing\section{0pt}{6pt plus 4pt minus 2pt}{0pt plus 2pt minus 2pt}
\titlespacing\subsection{0pt}{6pt plus 4pt minus 2pt}{0pt plus 2pt minus 2pt}
\titlespacing\subsubsection{0pt}{6pt plus 4pt minus 2pt}{0pt plus 2pt minus 2pt}

\newsavebox\abimagebox
\DeclareDocumentCommand \placeholder { o m }{%
    \sbox\abimagebox{\includegraphics[#1]{example-image-a}}%
    \begin{tikzpicture} 
    \node[draw, dashed, text width=\the\wd\abimagebox,minimum height=\the\ht\abimagebox, align=center, inner sep=0]{#2};%
    \end{tikzpicture}%
}%

\ifacm
    \acmDOI{}
    
    \acmISBN{}
    
    \acmConference[ACM Conference Template]{}
    \acmYear{}
    
    \acmPrice{}
\fi

\begin{document}
\pagenumbering{gobble}
\interfootnotelinepenalty=10000
\setlength{\belowdisplayskip}{3pt} \setlength{\belowdisplayshortskip}{3pt}
\setlength{\abovedisplayskip}{3pt} \setlength{\abovedisplayshortskip}{3pt}

\title{\paperTitle: Wireless Sensing and Localization of Contact Forces \\
on a Space Continuum}


\author{Agrim Gupta, C\'edric Girerd, Manideep Dunna, Qiming Zhang, Raghav Subbaraman, Tania K. Morimoto, Dinesh Bharadia\\
\{agg003,cgirerd,mdunna,qiz127,rsubbaraman,tkmorimoto,dineshb\}@eng.ucsd.edu\\
University of California, San Diego}

%

\ifacm

\begin{abstract}
\vspace{-5pt}
Contact force is a natural way for humans to interact with the physical world around us. However, most of our interactions with the digital world are largely based on a simple binary sense of touch (contact or no contact). Similarly, when interacting with robots to perform complex tasks, such as surgery, richer force information that includes both magnitude and contact location is important for task performance.
To address these challenges, we present the design and fabrication of \name
which is a `wireless' sensor, sentient to contact force magnitude and location. 
\name achieves this by transducing force magnitude and location, to phase changes of an incident RF signal of a backscattering tag.
The phase changes are thus modulated into the backscattered RF signal,
which enables measurement of force magnitude and contact location by inferring the phases of the reflected RF signal. \name's sensor is designed to support wide-band frequencies all the way up to 3~GHz. We evaluate the force sensing wirelessly in different environments, including through phantom tissue, and achieve force accuracy of 0.3~N and contact location accuracy of 0.6~mm.
\end{abstract}

\fi

\maketitle

\ifusenix

\begin{abstract}
\vspace{-5pt}
Contact force is a natural way for humans to interact with the physical world around us. However, most of our interactions with the digital world are largely based on a simple binary sense of touch (contact or no contact). Similarly, when interacting with robots to perform complex tasks, such as surgery, richer force information that includes both magnitude and contact location is important for task performance.
To address these challenges, we present the design and fabrication of \name
which is a `wireless' sensor, sentient to contact force magnitude and location. 
\name achieves this by transducing force magnitude and location, to phase changes of an incident RF signal of a backscattering tag.
The phase changes are thus modulated into the backscattered RF signal,
which enables measurement of force magnitude and contact location by inferring the phases of the reflected RF signal. \name's sensor is designed to support wide-band frequencies all the way up to 3~GHz. We evaluate the force sensing wirelessly in different environments, including through phantom tissue, and achieve force accuracy of 0.3~N and contact location accuracy of 0.6~mm.
\end{abstract}

\fi
\thispagestyle{empty}

\ifsurface
\section{Introduction}
\vspace{1mm}
\noindent Our sense of touch is critical for understanding and interacting with the world around us. 
While interacting with the physical world, force-sensitive mechanoreceptors in the skin respond to various vibrations, motions, pressures, and stretching of the skin to provide us with critical information on the location and magnitude of the stimuli~\cite{abraira2013sensory}.
Thus, if we want the next generation of tactile sensors to emulate how our skin reacts to stimuli, we need to both sense the magnitude and location of contact forces acting on the sensing surface.

Current skin-like continuum tactile sensors enable numerous critical applications. 
These applications mostly involve dexterous tasks to be performed via mechanical tools or robotic manipulators, rather than via human hands.
For example, in order to grasp and manipulate an object, a robot must be able to sense where and how firmly it is pressing the object~\cite{billard2019trends, deng2020grasping}. Another example can be seen during minimally invasive surgery, where a surgeon must operate inside the body with a surgical tool that naturally contacts numerous tissues throughout the procedure.
A sensing layer which acts like a skin covering the entire surgical tool could enable safer surgeries~\cite{al2020survey,dargahi2012tactile,koehn2015surgeons}, since the surgeon would know exactly where the tool is in contact with the tissues and with how much force.
In addition to these robotics applications, tactile sensing can supplement our interactions with the digital world.
Most of our current interactions with digital technologies occur with aid of a touchscreen, which binarizes human contact into simply touch/no-touch, and the richer information on contact force is typically lost. 
Augmenting our digital interfaces with the capability to sense the magnitude of the forces with which we interact with them could lead to more natural, intuitive, and realistic interactions, creating new possibilities for the evolving AR/VR settings~\cite{bai2020stretchable,zhu2020haptic,dangxiao2019haptic}.

Driven by these applications, design of such continuum sensor skins has been an active area of research over the past decade~\cite{icra2019,icra20201,icra20202,dahiya2013directions,soni2020soft,rosenberg2009unmousepad,zou2017novel,parzer2016flextiles,parzer2018resi,pointner2020knitted,lee2017durable}.
The common approach has been to create a sensing surface consisting of an array of discrete force sensitive resistors or electrodes, whose measurements are interpolated to reconstruct a continuum force profile.
However, this approach has prohibitive wiring costs~\cite{icra2019,icra20201,icra20202,dahiya2013directions,soni2020soft,rosenberg2009unmousepad,zou2017novel}, since it requires a wired link to obtain data from each individual sensor, as well as wires for satisfying the power requirements. 
In scenarios where space is a premium, including surgical robotic applications, this wiring challenge is exacerbated, and force sensing for the surgical robotics has been acknowledged as a `Grand-Challenge'~\cite{burgner2015continuum}.
One way to address the wiring requirements is to reduce the density of sensors in the surface and improve the interpolating algorithms~\cite{icra2019,icra20201,icra20202}.
A more drastic solution is to eliminate the wiring problem completely by creating new sensing modalities with modest power requirements such that both the sensor feedback and power can be delivered wirelessly~\cite{dahiya2013directions,soni2020soft,zou2017novel}.
 

Motivated by these challenges, we present \paperTitle, which makes progress 
in this direction by sensing force magnitude and location over a 1-D continuum by leveraging backscattering techniques. 
Rather than generating a wireless signal to feed back the sensor readings, which would require power-hungry electronics, \name's sensor transduces force magnitude and location directly onto the reflections of the incident RF signals. The design has very minimal power requirements, and consists of only one antenna, a small identification unit, and a force continuum surface.
Thus, \paperTitle presents a new tactile sensing modality, which makes headway towards batteryless wire-free sensor skins.

\begin{figure}[t!]
\centering
  \includegraphics[width=\linewidth]{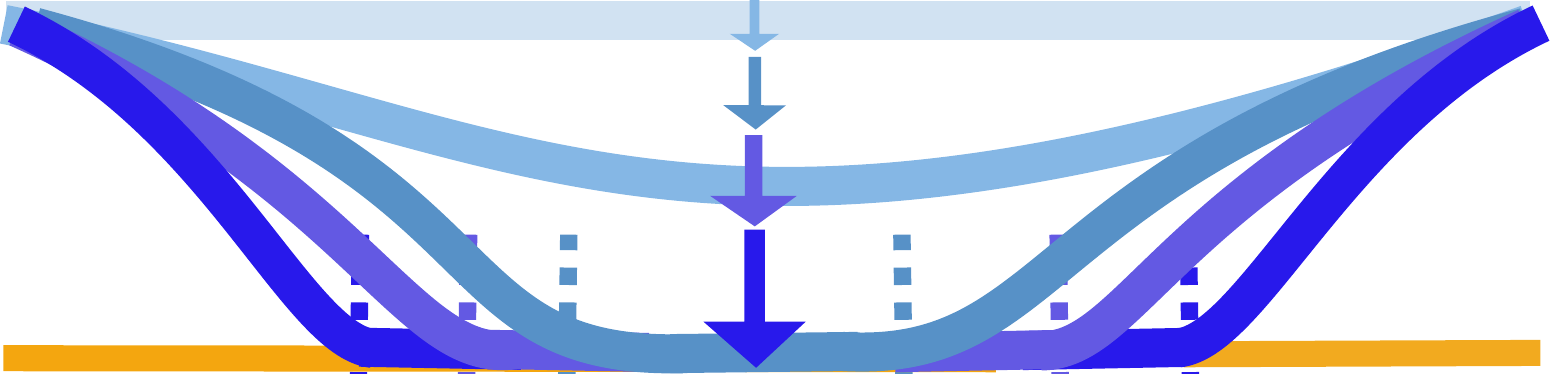}
  \caption{Beam bending in effect of contact force:
  As contact force increases (shown via increasing arrow lengths), the top beam bends and collapses more and more onto the bottom beam}
  \label{fig:intro_a}
\end{figure}

The key enabler for such a low-powered design is the transduction mechanism, which modulates the reflected signal with information on the contact force and its location, by altering the RF signal parameters as applied force on the sensor changes.
To achieve this, \name links contact force, a mechanical entity, to RF signal parameters by combining classical beam bending models and RF transmission line concepts, using a novel sensor surface. 
This sensor surface consists of two parallel conductive traces, similar to a microstrip line, augmented with a soft specialized polymer beam.
As a force is applied at a specific location on the sensor surface, the beam bends, causing the traces to connect (Fig. ~\ref{fig:intro_a}). 
From the RF perspective, this beam bending leads to shorting of the traces, which causes reflection of signals. 
From the mechanical perspective, the soft beam allows us to use beam bending models to characterize how the shorting phenomenon changes as the applied force increases.


Essentially, the shorting points shift towards the ends of the sensor, as the applied force increases and the soft layer of the beam bends and flattens on the bottom trace (Fig.~\ref{fig:intro_a}).
By estimating the shorting lengths from both ends of the sensor, we can determine the magnitude and location of the applied force.
The shorting lengths are related to the signal phases measured on both the ends of the sensor. 
Basically, the longer the signal travels on the sensor surface, the more phase change it will accumulate.
The goal at the wireless reader is to measure the accumulated phases due to signal propagation on the sensor surface from both the ends, in order to use the transduction mechanism to sense and localize the forces.

To enable sensing of these phases by the wireless reader, the phases from both ends have to be disambiguated, and thus each end has to be given an identity. To do so, a naive solution would be to have RF switches toggling on-off with different frequencies on either ends ($f_{s_1},f_{s_2}$, Fig.~\ref{fig:intro_b}), with the toggling frequency providing the unique identity to each of the ends.
However, this naive solution does not work out of the box, because the two ends are electrically connected to each other via the transmission line, causing signals to leak from one end to the other, which would in turn cause intermodulation effects. To resolve this issue, \name comes up with a creative RF switch toggling strategy, which not only provides electrical isolation to combat intermodulation, but also provides different identities to these ends in terms of different frequency shifts. 
Thus, the external wireless reader is able to view the sensor ends as having different identities in frequency domain, as envisioned by the intuitive scheme in Fig.~\ref{fig:intro_b}, with the intermodulation problem abstracted out via the intelligent toggling scheme.

\begin{figure}[t!]
  \centering
  \includegraphics[width=0.9\linewidth]{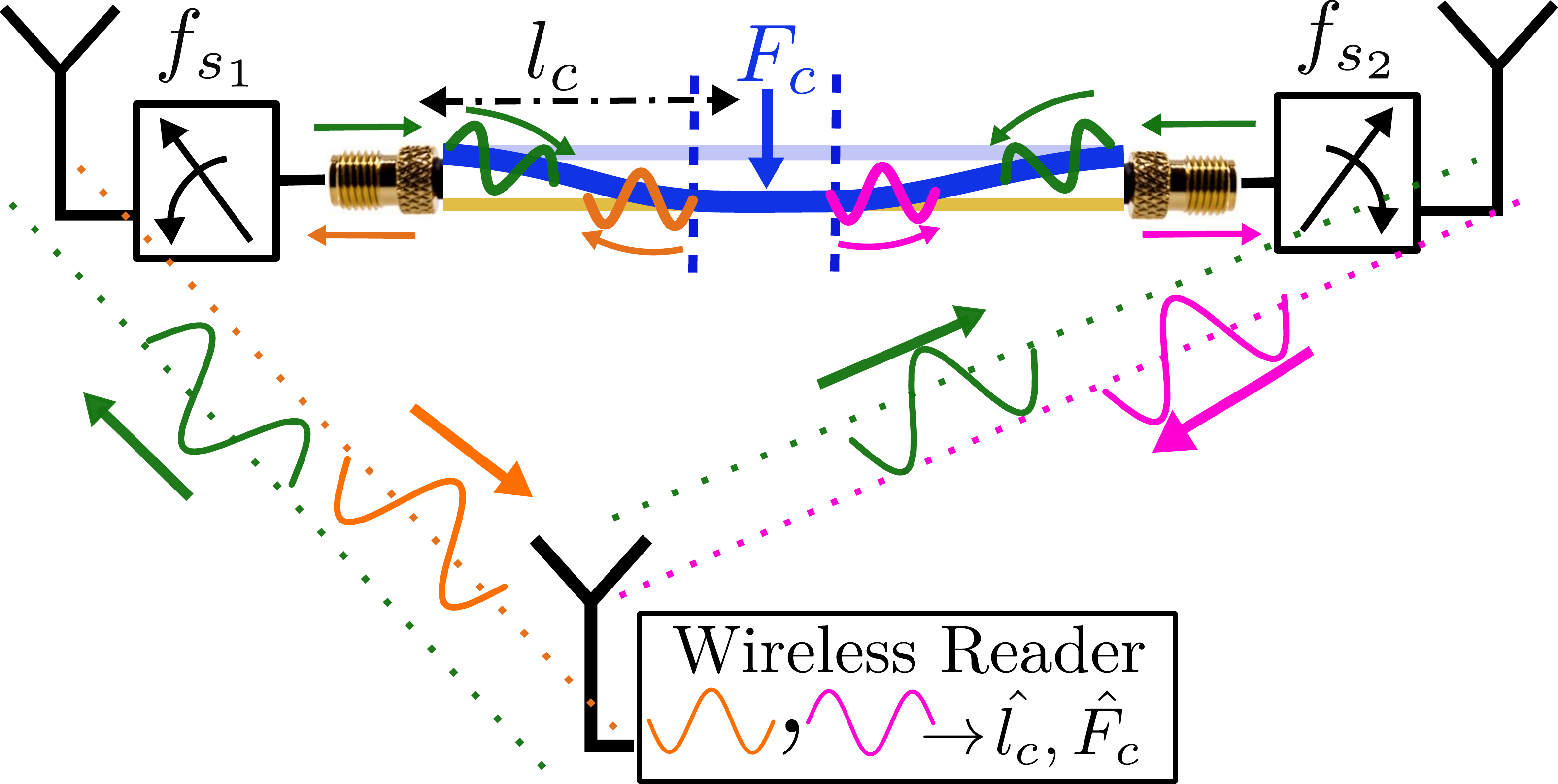}
  \caption{
    The key insight of \name is to view the parallel beams as a microstrip line.
    Force $F_c$ and it's location $l_c$ gets transduced onto changes in the reflections due to the line shorting caused by beam bending. The wireless reader uses the reflected signals to estimate $l_c, F_c$
  }
  \label{fig:intro_b}
\end{figure}


The final piece in \name is designing the wireless reader, such that it can use any wireless device (like WiFi (OFDM) or LoRa (FMCW)) with wide-band transmission to read the \name's force and location. The task of the wireless reader is two-fold, first identify and isolate the signals coming from the sensor and second, accurately track the phase of the sensor signal. Since the wireless phase observed by the reader can also be altered by various entities in the environment, the task of reading phase changes stemming only from the sensor is non-trivial. Hence, \name designs a novel signal processing algorithm which utilizes periodic wide-band channel estimates to pick up the reflection signatures from the sensor to isolate the signal, as well as to read the phase changes at multiple frequencies, providing robustness to the phase sensing requirements for the proposed force transduction mechanism.

We designed and fabricated the sensor with a soft-polymer augmented microstrip line, which is `force-sensitive'. That is, the microstrip line sensing surface has bending properties which maximize the phase changes transduced by contact forces. This sensing surface was retrofitted with RF-switches and antenna to enable backscattered feedback. 
The fabricated sensor works for the entire sub-3 GHz~verified with the test equipment (vector network analyzer). We evaluated the \paperTitle sensor abilities to report force magnitude and location in multiple settings, both indoors and inside a body-like environment using gelatin. We used USRP radios as the readers, and tested the sensor at 900~MHz and 2.4~GHz, which are the two most popular ISM bands. We show that the sensor can be read up to 5 meters of range over the air, and show the algorithm working even with propagation through the gelatin-based muscle/fat/skin tissue layers composition similar to the human body to demonstrate the surgical applications.
We achieved phase sensing with an accuracy as low as $0.5^{o}$, giving us a force resolution of $0.3$~N, and location accuracy of $0.3$~mm.  
We also showcase the ability to read from multiple sensors, by sensing forces from 2 sensors simultaneously.
Finally, we even evaluated our force sensor with a user pressing with his hand, and we achieved force resolution of $0.3$~N, and location accuracy of $0.3$~mm. In fact, recent interfaces for Human-Computer Interactions (HCI) work shows that similar resolution ($0.2$~N) is required to support force enabled gestures on smartphones and desktop computers~\cite{antoine2017forceedge}. We believe this is the first step towards enabling numerous force sensing applications.

\fi

\ifshowsumm
\else
\section{Background and Motivation}
\vspace{1mm}
\noindent The problem of sensing tactile phenomenon over a surface continuum has attracted considerable research interest~\cite{rosenberg2009unmousepad,lee2017durable,icra2019,icra20201,icra20202,parzer2018resi}. 
The usual approach has been to densely populate the surface with either force sensitive resistors~\cite{rosenberg2009unmousepad}, electrodes~\cite{lee2017durable,icra2019,icra20201,icra20202}, or force sensitive yarns~\cite{parzer2018resi}.
The continuum sensing is performed by interpolating over the sensor readings of these discrete sensors.
Numerous papers in the past decade have raised the issues stemming from the wiring requirements of the developed sensor skin modalities~\cite{dahiya2013directions,soni2020soft,zou2017novel}. 

Researchers have tried addressing these issues by considering sparser deployments, such as considering sensors only on the boundaries~\cite{lee2017durable}, or populating the sensors in a minimal way across the surface~\cite{icra2019,silvera2014electrical}. Although these efforts have reduced the wiring requirements for sensing considerably, these surfaces still lack a solution to both feedback the sensor readings wirelessly, as well as get rid of wired battery connections required for these sensing efforts. Recent review papers have advocated the need of powering up these sensing surfaces with energy harvesting methods to alleviate the battery requirements~\cite{soni2020soft,zou2017novel}, and a backscatter-enabled sensor is a promising approach to address the battery concerns. 

Before re-designing the sensing modalities to be compatible with low-powered backscatter communications, a key question to answer is whether a hybrid solution would work. That is, can we take one of the sensing solutions requiring the least number of wired connections across the surface~\cite{lee2017durable,icra2019} and feedback the sensor readings via currently developed backscattering RFICs\footnote{This fusion of sensor skin + backscatter RFIC has not been yet demonstrated, however we consider it as an hypothetical scenario}~\cite{wang202020,hitchhike}.
However, this solution won't suffice since these backscatter links typically work with a RF energy harvestor, which generates small voltages capable of powering a small RFID chip, and not a large continuum sensing surface.
Further we would need to sense multiple voltages from these electrodes via an array of ADCs (Fig.~\ref{fig:sec3_fig}), managed by a micro-controller, which would then digitize and buffer the data for transmission through the low capacity backscatter link. 
\begin{figure}[t!]
\begin{subfigure}{\linewidth}
  \centering
  \includegraphics[width=\linewidth]{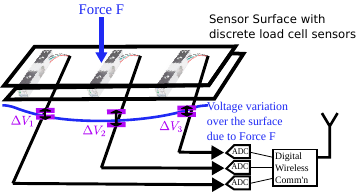}
  \caption{Shows a possible continuum sensing approach with existing wired force sensors, which sense discrete voltage changes over a continuum ($\Delta V_1,\Delta V_2,\Delta V_3$), and how this sensing approach could be made wireless}
  \label{fig:wichlorian_vs_trad_a}
\end{subfigure}
\begin{subfigure}{\linewidth}
  \centering
  \includegraphics[width=0.9\linewidth]{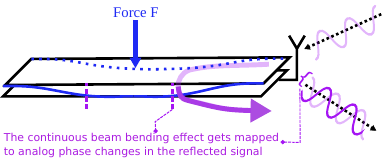}
  \caption{\name transduces the continuum force information directly onto analog backscattered phase changes without discretization}
  \label{fig:wichlorian_vs_trad_b}
\end{subfigure}
    \caption{Force feedback design, \paperTitle in comparison with a possible wireless extension to existing sensing modalities}
    \label{fig:sec3_fig}
\end{figure}

Hence, \paperTitle attempts a RF-only analog approach, where the sensing modality directly transduces force and its location into wireless signal phase changes, which can be read by a radio over the air. 
The argument here is that, if analog phase readings can be fed back accurately, it would require much less power than procuring the analog readings, digitizing/buffering them, and then sending them over the backscatter link.
Thus, the novel force to phase transduction mechanism, coupled with the analog phase feedback, fulfills both the key requirements for low powered tactile sensing -- the ability to sense over a continuum and low-powered battery-free operation.

\section{Design of \paperTitle}

\noindent In this section, we present the design principles of \name. First, we describe \name's force transduction mechanism, which translates the force and its application location to changes in the RF signal phase. Next, we present novel algorithms to measure changes in the RF properties to deduce the force and its application location wirelessly. Finally, we conclude with a robust channel measurement technique that uses a wireless waveform to read the sensors while rejecting multi-path.

\subsection{Force Transduction Mechanism}
\label{subsec:force_transduction}
\noindent As a first step towards a backscattered force sensor, \name has to formulate a transduction mechanism which relates force magnitude and application location, to parameters like RF signal amplitude and phase, which can then be used to modulate the sensor readings onto the reflected signals. 
The challenge here is to take an object (like transmission lines) which supports RF signal propagation, and make it force-sensitive. That is, RF propagation in this object should give significant changes in its signal parameters as we press the object at different locations with varying force magnitudes. 
In this section, we will elaborate on how \name makes microstrip lines `force-sensitive'. 


\begin{figure}[t!]
\centering%
\includegraphics[width=\linewidth]{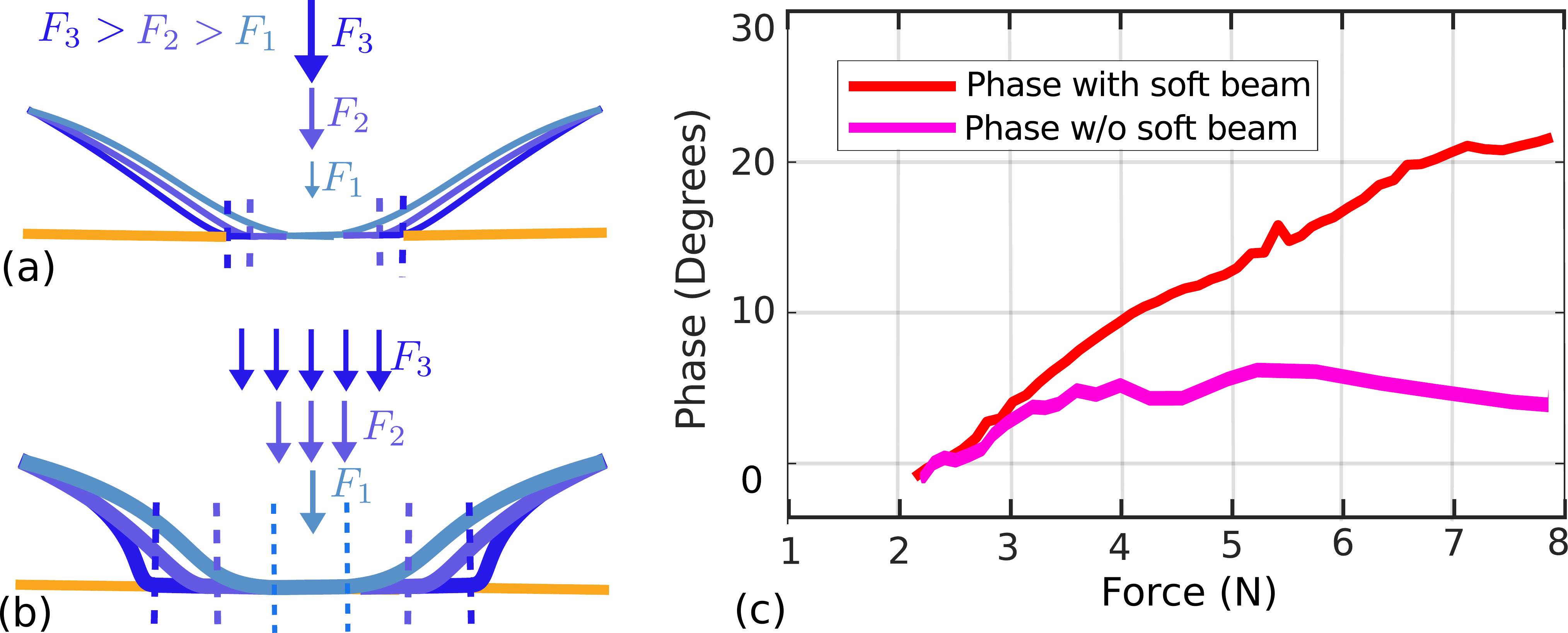}%
 \caption{Bending of (a) Thin, (b) Soft beam augmented thick trace, as forces increase ($F_1<F_2<F_3$).
 The soft beam distributes forces along the length, which leads to profound phase changes (c) as compared to thin traces w/o soft beam}%
\label{fig:mech_impl}
\end{figure}

A microstrip line traditionally consists of two parallel conducting traces-- the signal trace and the ground trace. 
A force applied to the microstrip line would cause the traces to bend and come in contact with each other, which shorts the line and leads to signal reflections. 
The reflections produced by this shorting have different phase accumulation based on the location of pressing. However, this reflected phase
is not sensitive to force magnitude at all. That is, irrespective of the contact force applied, the traces will short each other only in the vicinity of a single point (Fig.~\ref{fig:mech_impl}a). The contact point invariance leads to a near invariant phase response as force is changed (Fig.~\ref{fig:mech_impl}c), therefore preventing the measurement of force through phase changes.

\paperTitle modifies the traditional microstrip line by augmenting a new soft, flexible beam on top of the signal trace to address this problem and make the microstrip line force-sensitive. The key insight here is that the soft beam distributes the force along the length of the trace 
(Please refer to~\cite{mechPaper} for details on the mechanical implementation of the sensor). The distributed force leads to a finite length of the signal trace to come in contact with the ground trace, creating two distinct shorting points, as shown in Fig.~\ref{fig:mech_impl}b. Further, these shorting points shift towards the ends as the applied force magnitude increases. Varying shifts induce different phase changes since the signals travel less distance on the microstrip line before getting reflected at the shorting points. Hence, the reflections caused by higher magnitude forces accumulate less phase relative to the reflected phases when lower force was applied. Thus, the soft beam augmented microstrip line allows phase to force transduction (Fig.~\ref{fig:mech_impl}c).

\noindent This beam bending effect manifests itself in the form of a varying phase-force relationship depending on the contact force's point of application (Fig.~\ref{fig:transduction}, top images). A force applied in the middle of the sensor compresses it symmetrically, and therefore the reflected signals from both the ends show similar phase changes. In contrast, a force that acts asymmetrically will disproportionately compress the smaller length of the beam. Therefore, the end near the smaller length shows a higher phase shift than the end near the longer one. The longer length collapses onto the bottom trace, leading to an almost stationary shorting point as the force increases (Fig.~\ref{fig:transduction}, bottom images). These varying phase changes  that depend on the location of a contact force also allow \paperTitle to localize the force application point. Thus, the double-ended measurement allows us to estimate the applied force's magnitude and its application location along the sensor length. However, at the same time, this asymmetric behavior of phase change with force contact location necessitates sensing phases from both ends of the sensor. 

\begin{figure}[t!]
  \includegraphics[width=\linewidth]{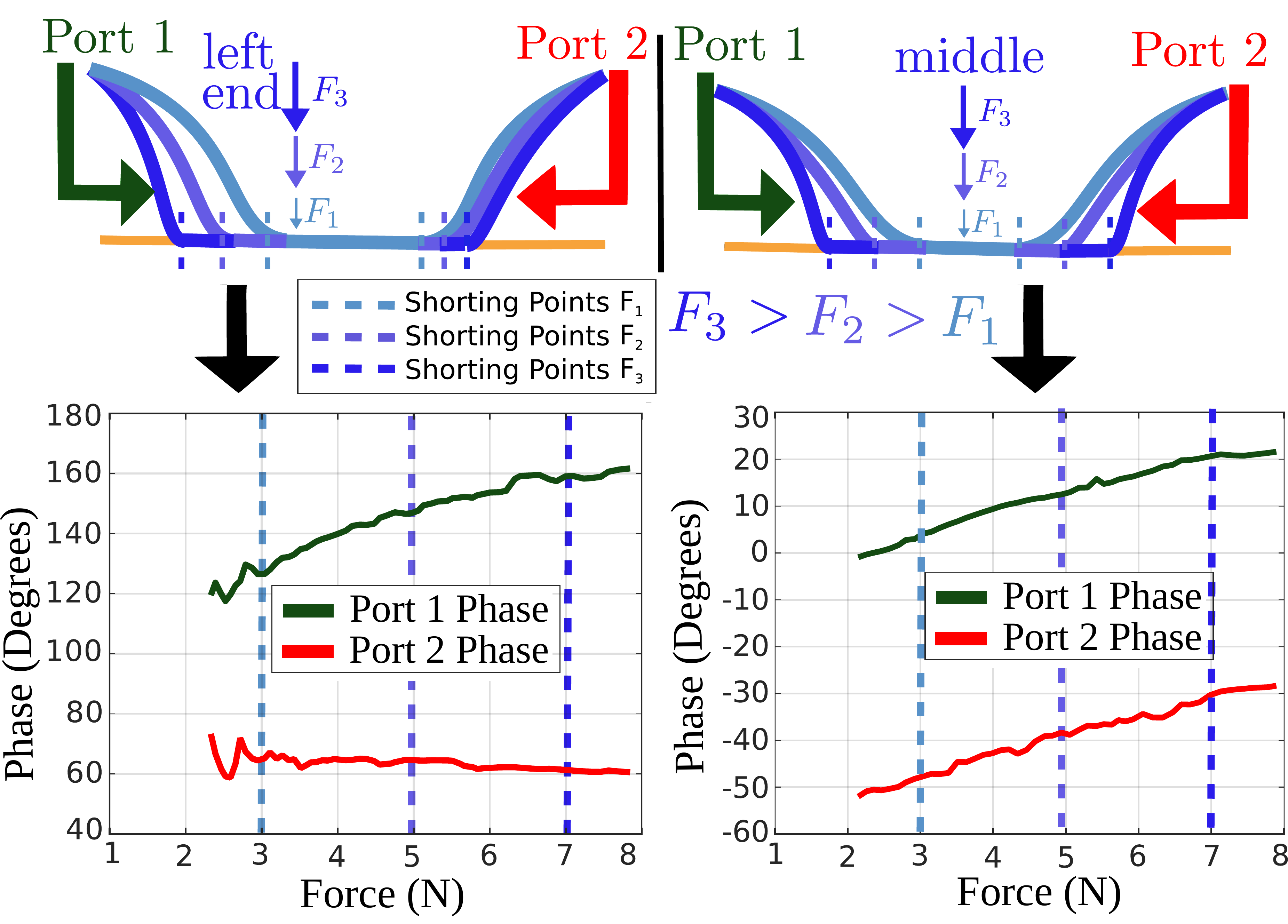}
\caption{The shorting points shift due to increased force because of bending of the soft beam, changing the reflected phase. When the force is applied at the middle, we observe symmetric phase change across the two ports, whereas when force is applied towards the ends, we see asymmetric phase changes, due to the beam bending mechanism illustrated in the top row}
\label{fig:transduction}
\end{figure}

\subsection{Two-ended backscatter modulation}
\label{Subsec:double-ended}
\noindent As described in the previous section, sensing phases from both ends of the sensor forms the cornerstone of the phase to force transduction mechanism. 
This is because it allows to disambiguate the different force profiles observed as the sensor is pressed at different locations, which basically allows us to both locate and then measure how much force was being applied.
In this section, we will go over how to attempt this double ended phase sensing via wireless backscatter sensing.

The first and foremost thing which any backscatter sensing solution needs, is an ability to give an identification to the reflections occurring at the sensor. 
This identification helps the wireless reader isolate the sensor reflections from the environmental clutter. 
A popular technique to do so has been to use RF switches toggling at different frequencies as identification unit~\cite{luo20193d,josephson2019rf,freerider}. This technique basically multiplies the incident signal with an on-off modulation of certain switching frequency. 

In frequency domain, this operation leads to frequency shifts corresponding to the switching frequency. Putting this mathematically, say the sensor receives the excitation signal $s(t)$ and reflects $s(t)m(t)$ where $m(t)$ is a square wave, with time period $T$. Expanding $m(t)$'s Fourier series, we get odd harmonics, $m(t) = \sum_{k\in (2i+1), i\in \mathbb{Z}} \frac{1}{|k|}e^{(j2\pi kf_st)}$ where $f_s=\frac{1}{T}$. 
Ignoring the high order harmonics, we get reflected signal as $r(t) = s(t)m(t) \approx s(t)e^{j2\pi f_s t}$. 
Hence the reflected signal will be shifted by the switching frequency $\pm f_s$, which allows the reflected signal, $r(t)$, to be isolated from the excitation signal, $s(t)$ in the frequency domain. 

A naive extension of this idea to the double ended sensing problem at hand, would be to have switches at both ends of the sensor and make them switch at different frequencies ($f_{s_1}/f_{s_2}$). 
Theoretically, this solution should give separate identities to reflections emanating from both the ends.
However, the problem at hand is inherently coupled to allow for such naive de-coupled solutions, because both the ends are physically connected to each other via a microstrip line.
When both the switches are toggled-on, the signals will propagate through the sensor and leak out from the other end (Fig.~\ref{fig:intermod}).
This causes intermodulated reflections, where the reflected signal would be partially modulated by both toggling frequencies, leading to muddled up identities. 

\begin{figure}[!t]
\centering
\includegraphics[width=0.8\linewidth]{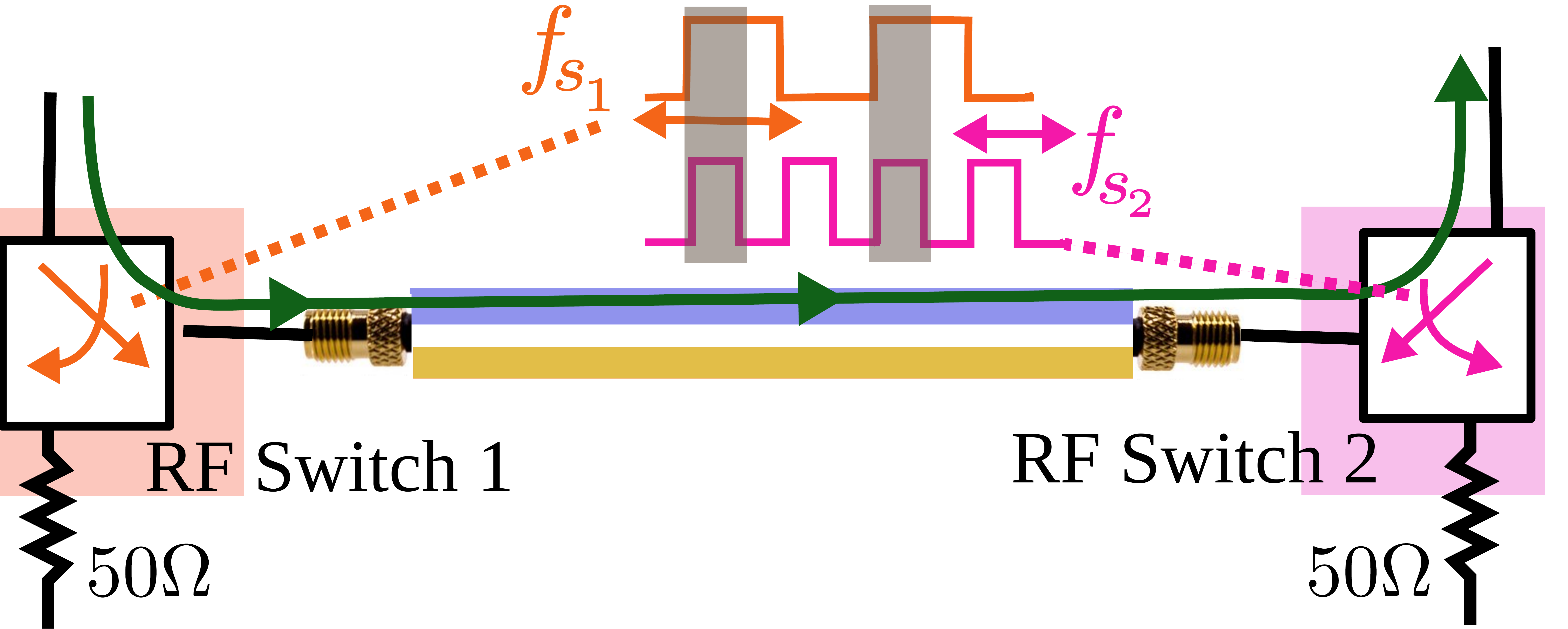}
\caption{RF switches toggle between sensor (on) and 50 $\Omega$ (off) depending on the control input. Different freq. clocks as control inputs introduce intermodulation due to both switches being toggled on at the same time (grey shaded time instants)
}
\vspace{0.3mm}
\label{fig:intermod}
\end{figure}
The challenge in avoiding the intermodulation effects is that the sensor has to reflect signals from the opposite end when a switch is toggled on. Using reflective RF switches
in the off state allows us to make either of the ends reflective, under the restriction that the other should be off when one switch is on.
Said differently, we want to design a coupled two-ended switching scheme, which gives separation in frequency domain, under the constraint that both switches are not `on' at the same time. 
The unique insight which allows \paperTitle to have such an switching scheme, is the use of duty cycle properties of square wave Fourier series. 

In a standard square wave, with $50\%$ duty cycle, all the even harmonics (i.e. every second harmonic) are absent. Similarly, in a wave with $25\%$ duty cycle, every fourth harmonic would be absent. Hence, a frequency $f_s$, $25\%$ duty cycle square wave will give modulation at $\underline{\mathbf{f_s}},2f_s,3f_s,\cancel{4f_s} 5f_s, \ldots$  Similarly, a frequency $2f_s$, $25\%$ duty cycle square wave will give modulation at $2f_s,\underline{\mathbf{4f_s}},6f_s,\cancel{8f_s},10f_s\ldots$ Observe that a combination of these 2 clocks will cause interference at $2f_s$, but can be read up individually at $f_s$ for the former clock, and $4f_s$ for the latter clock. Hence, a combination of these 2 clocks can provide separation in the frequency domain. Also, by controlling the initial phases of these two clocks, we can suppress the intermodulation problem as well. This is possible because when one clock is high, the other clock will be guaranteed to be low and vice versa (Fig.~\ref{fig:two-sided}). Hence, at any given time, only one port will be on, and other port will be reflective open. 

Further, this clocking design allows us to reduce the form factor requirements, instead of having $2$ antennas, one for each end of the sensor, we can just have just a one antenna design using a splitter. Since the clocking strategy provides separation in the frequency domain, we can add the modulated signals from the either ends via a splitter. Thus, the wireless reader can identify the two ends by reading at $f_s, 4f_s$ frequency shifts.
\begin{figure}[t!]
\centering
\includegraphics[width=\linewidth]{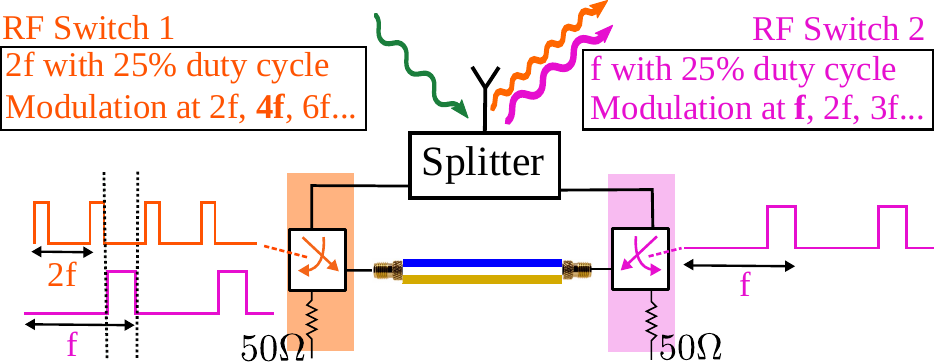}
\caption{The duty-cycled modulation ensures that switches aren't toggled on at once, as well as providing freq. separation}
\label{fig:two-sided}
\vspace{0.3mm}
\end{figure}

\subsection{Sensing Forces at the Wireless Reader}
\label{Subsec:wideband}
\noindent Till now, we have described the phase-force transduction mechanism, and delineated a method to give disambiguated identities to both ends of the sensor. Now, in this section, we move on to the description of how the wireless reader is designed. 
We design our wireless reader to detect the separate identities stemming from frequency shifts, and then extract the valuable phase information which allows us to sense and localize forces.
The key insight of \paperTitle here is to view the frequency shifts from the sensor as `artificial-doppler' and use wide-band channel estimates in order to estimate the doppler and thus isolate the signal coming from the sensor. 
This approach to view the backscatter tag's frequency shift as an artificial doppler has been also utilized in some of the recent past work~\cite{josephson2019rf}. Finally, to obtain the required analog phase estimates required to sense and localize the forces, we utilize that fact that force, a mechanical quantity, changes slowly (at about 1 kHz rate~\cite{maciel2009using,wagner2002role,wang2018toward}), as compared to MHz's of RF bandwidth.
This allows us to group the channel estimates and perform a `short-time phase transform', which enables us to track the phase shifts at the two artificial doppler bins corresponding to the two ends of the sensor.

The algorithm \name uses to extract the backscattered phases embedded inside the wideband channel estimates is visually illustrated in Fig.~\ref{fig:ofdm_expl}. Say we are estimating the channel periodically after every $T$ seconds, with frequency steps of $F$. 
If we use OFDM channel sounding strategy, $T$ will be the time of the OFDM frame, and $F$ will be the subcarrier spacing.
We denote $H(kF,nT) = H[k,n]$ where $k$ is subcarrier index and $n$ is time index. 
If there are $M$ multipaths in addition to the signal coming from the sensor, we can write the channel estimates from geometric channel model as 
\begin{equation}
\begin{split}
H[k,n] &= \sum_{i=1}^M \alpha_i e^{-j2\pi k F \frac{d_i}{c}} + (s_1(nT) e^{-j\phi^1_n} \\[-5pt]
&+s_2(nT)e^{-j\phi^2_n})\alpha_s e^{-j2\pi k F \frac{d_s}{c}}
\label{eqn:chan_est}
\end{split}
\end{equation}

\noindent Here, $\alpha_i$ is the attenuation factor for the $i$-th path, $d_i$ is the distance separation between the TX-reflector and reflector-RX, and $\phi^1_n, \phi^2_n$ is the phase accumulated from the RF propagation in the microstrip line sensor at time index $n$ from sensor end $1$ and sensor end $2$. $s_1(t),s_2(t)$ are the duty-cycle square wave modulations to give intermodulation free frequency identities at $f_s,4f_s$ as discussed in Section~\ref{Subsec:double-ended}.
Ignoring the higher harmonics terms in $s_1(t),s_2(t)$, we get 
\begin{equation}
\begin{split}
H[k,n] &\approx \sum_{i=1}^M \alpha_i e^{-j2\pi k F \frac{d_i}{c}} + (e^{j2\pi (f_{s})nT}e^{-j\phi^1_n}\\&+e^{j2\pi (4f_{s})nT}e^{-j\phi^2_n})\alpha_s e^{-j2\pi k F \frac{d_s}{c}}
\label{eqn:chan_est2}
\end{split}
\end{equation}

Now, to isolate the signal from the sensor, we take FFT over the $n$ index, to obtain $\tilde{H}[k,f]$. We observe $N$ channel snapshots to calculate, $\tilde{H}[k,f] = \sum_{n=1}^N H[k,n] e^{-j2\pi f n T}$. 
Assuming $\phi^1_n, \phi^2_n$ stay constant over the period of $N$ snapshots, at $f_s$, $4f_s$, we have,
\begin{equation}
\begin{split}
\tilde{H}[k,\{f_s,4f_s\}] &= \sum_{n=1}^N H[k,n] e^{-j2\pi \{f_s,4f_s\} n T}\\[-5pt]
&= \alpha_s e^{-j2\pi k F \frac{d_s}{c}} e^{j\phi^{\{1,2\}}_n}
\label{eqn:sensharmonics}
\end{split}
\end{equation}

Observe that for this transform, the nyquist frequency would be $\frac{1}{2T}$, and hence, $f_s$ has to be chosen such that $4f_s \leq \frac{1}{2T}$. 
The switching frequency $f_s$ can be related to an equivalent Doppler, $f_s = \frac{f_c v}{c}$, and thus an object in the environment moving at velocity $v = \frac{c f_s}{f_c}$ would create interference with the sensor signal. However, the chosen $f_s$ is large enough so that this equivalent speed is so high to guarantee that, the signal observed in the frequency bins corresponding to $f_s, 4f_s$ are free from multipath clutter.

However, recall that while writing Eqn.~\eqref{eqn:sensharmonics}, we assumed $\phi^1_n, \phi^2_n$ stay constant as $n$ goes from 1 to $N$. That is, the transform is only valid when the phases from the sensor ends do not change much over the period of taking the transform. This is a reasonable assumption to make, since the sampling is occurring in MHz rate, whereas force will at max change in rate of kHz, since it is a mechanical quantity.
However, we can not obviously assume the phase to stay constant forever, and, the phase will change as we apply force on the sensor which would move the shorting points. 
More importantly, we need to not only tweak the standard doppler transform to respect phase change, we also need to estimate the phase changes in order to estimate the forces.
\begin{figure}[t!]
\centering
\includegraphics[width=\linewidth]{figures/fft_ofdmalgo.pdf}
\caption{\paperTitle's reader utilizes wideband channel estimates to isolate sensor signal from multipath in doppler domain. Arranging the channels into `groups' allows to read phase changes across subcarriers to give robust measurements}
\label{fig:ofdm_expl}
\end{figure}
Thus, \paperTitle designs an algorithm similar to the familiar short-time transforms. The algorithm divides the channel estimates into groups of $N_g$, referred to as `phase-groups'. For each phase-group we first take the harmonics FFT as described earlier and obtain two $K\times1$ vectors from Eqn.\ref{eqn:sensharmonics} for FFT frequency $f_s$, $4f_s$. Assume that there are $G$ such phase groups, i.e. $N=GN_g$. For all the $N_g$ samples of $g$-th group, $\phi^1_n,\phi^2_n \approx \phi^1_g \phi^2_g \forall n \in \{1,2,\ldots N_g\}$ from the choice of $N_g$ to respect the time it takes for the force to become effective. The output at $g$-th phase group, $k$-th subcarrier, after harmonics FFT at $f_s, 4f_s$, is denoted as $P_1[k,g],P_2[k,g]$. 
\begin{equation}
    P_{\{1,2\}}[k,g] =\tilde{H}[k,\{f_s,4f_s\}] = \alpha_s e^{-j2\pi k F \frac{d_s}{c}} e^{j\phi^{\{1,2\}}_g}
\end{equation}
To get rid of the air phases, we can obtain the phase change between 2 groups by conjugate multiplication:
\begin{equation}
\begin{split}
\tilde{P}_{\{1,2\}}[k,g] &= P_{\{1,2\}}[k,g+1]*\text{conj}(P_{\{1,2\}}[k,g]) \\[-5pt]
          & = \alpha_s^2 e^{j(\phi^{\{1,2\}}_{g+1}-\phi^{\{1,2\}}_g)}
\end{split}
\end{equation}
Hence, we have
\begin{equation}
\angle{\tilde{P}_1[k,g]} = \phi^1_{g+1}-\phi^1_g, \angle{\tilde{P}_2[k,g]} = \phi^2_{g+1}-\phi^2_g 
\end{equation}
for each subcarrier $k$. Observe that the right side of the equation is the phase change independent of $k$, which entails that we have $K$ independent estimates of the phase change from the $K$ subcarriers. Thus, we can estimate very precise phase changes by averaging over these $K$ independent estimates, allowing \name to calculate very precisely the analog phase changes.

\begin{figure}[t!]
    \centering
    \includegraphics[width=\linewidth]{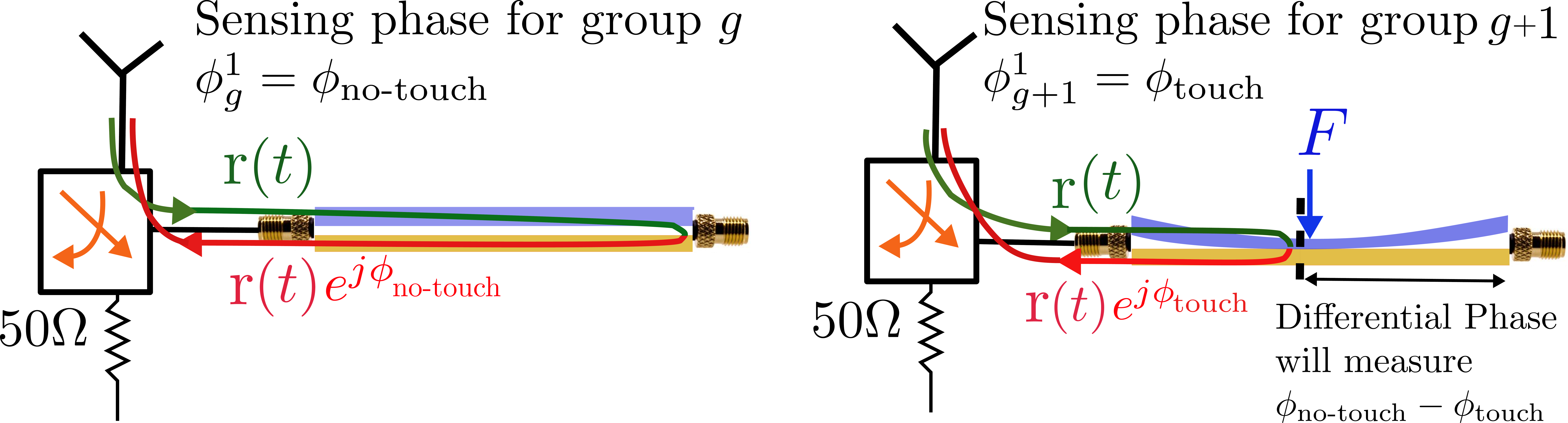}
    \caption{Differential Phase measurements can be compensated with fixed quantity $\phi_{\text{no-touch}}$ procured via calibration, to obtain the quantity of interest $\phi_{\text{touch}}$, which varies with force magnitude and location}
    \label{fig:diff_phase}
\end{figure}

The last piece in the puzzle to conclude the design section, is to internalize how can we use differential phase to sense and localize the contact forces. In fact, since force is an event based quantity, that is, unlike quantities like temperature, moisture, which sense the ambient quantity, tactile sensors have to sense the force from an `touch-event' which exerts certain force on the sensor at a given location. When we measure the differential phase between the `no-touch' and `touch' events, we can measure differential phase and obtain the absolute phase by simply subtracting the phase which the waves accumulate when the sensor is at rest (Fig.~\ref{fig:diff_phase}). Since this no-touch phase is a fixed quantity and depends only on the length of the trace, we measure it beforehand via a VNA setup and compensate. Hence, compensating the differential phase with the VNA calibrated no-touch phase allows us to recover phases from both the ends, which can then used by the transduction mechanism in order for estimation of force magnitude and location at the reader.

\section{Implementation}

\subsection{Microstrip line RF Interfacing}
\noindent To support the two broad applications targeted by \paperTitle, the sensor must give good RF propagation performances at 900 MHz (for in-body sensing applications) and at 2.4 GHz (to be compatible with Wi-Fi/Bluetooth standards). From our simulations (Section~\ref{section:hfss_sims}), we expect that by having the ratio between trace width and sensor height to be around $4:1$, we should get good impedance matching. Hence, we design an air substrate microstrip line with trace width of 2.5~mm, ground trace width of 6~mm and height of 0.63~mm, for a sensor length of 80~mm. To verify the impedance matching, we assess the RF design performance of the sensor in terms of insertion/thru losses. For this, we perform a 2 port amplitude/phase analysis using VNA. As visible in Fig.~\ref{fig:s11_mag_plot}, this leads to a S11/S22 ratio below -10~dB over the entire frequency range from 0 to 3~GHz, along with linear S12 phase, which justifies the broadband nature of the sensor.


\subsection{Forming the sensor model with soft beam microstrip line}
\label{subsec:vna_meas}


\noindent After having verified the RF properties of the microstrip line, we fabricate the soft layer using Ecoflex 00-30 (a commonly used elastic material~\cite{park2012design,elsayed2014finite,teng2019soft}). The Ecoflex layer is 
placed onto the top trace to create the \name sensing surface with thick traces, endowed with the novel phase to force transduction mechanism. 
\begin{figure}[t]
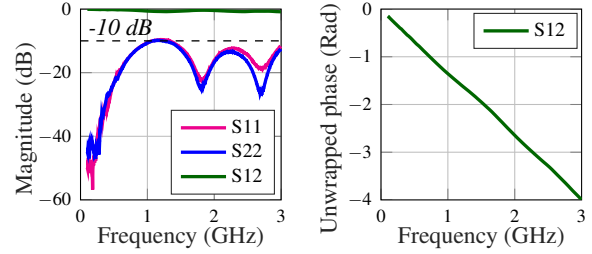

\begin{subfigure}[t]{0.49\linewidth}
 {\input{results/s11-s33-mag}\label{fig:s11_mag}}
\put(-78,84){\small{\textit{\shortstack[l]{-10~dB}}}}
\end{subfigure}
\hfill
\begin{subfigure}[t]{0.49\linewidth}
 {\input{results/s11-s33-phase}\label{fig:s11_phase}}
\end{subfigure}
\caption{2 port RF profiling of the sensor. S11 stays below -10~dB across 0-3 GHz, with S12 around 0~dB with linear phase.}
\label{fig:s11_mag_plot}
\end{figure}

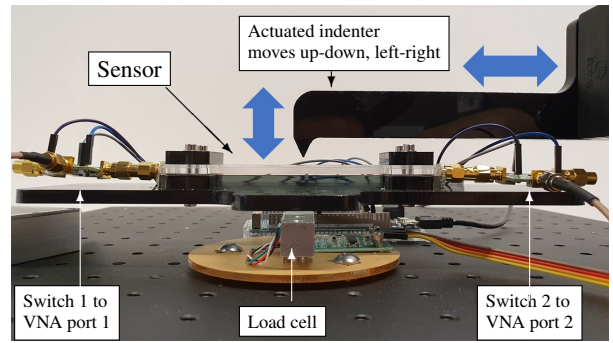
\begin{figure}[b!]
  \centering
  \includegraphics[width=\linewidth]{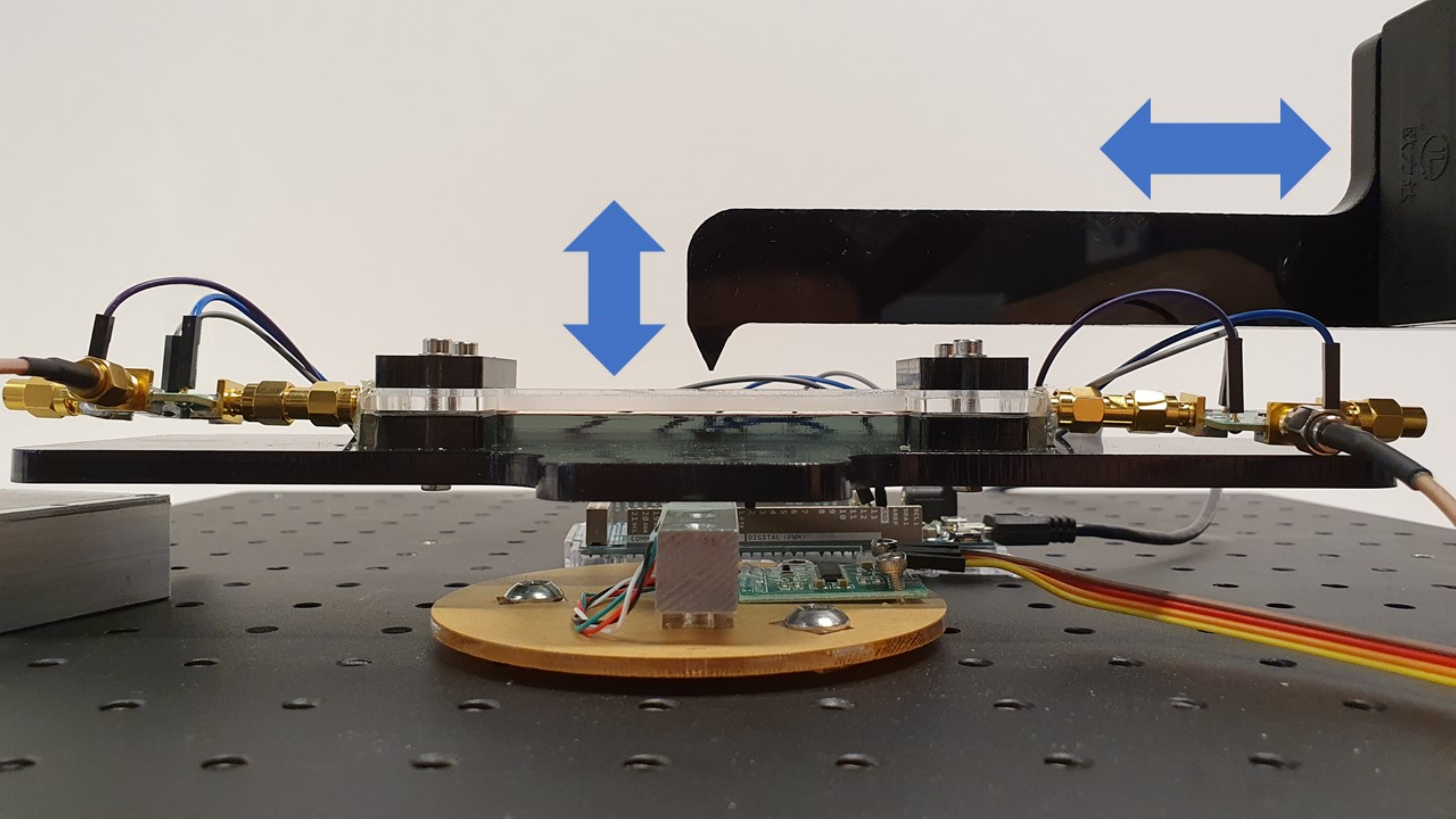}
  \put(-197,100){\fcolorbox{black}{white}{\small{\shortstack[l]{Sensor}}}}
  \put(-167.5,96){\vector(1,-1){25}}
  \put(-140,107){\fcolorbox{black}{white}{\scriptsize{\shortstack[l]{Actuated indenter\\
  moves up-down, left-right}}}}
  \put(-105,102){\vector(0,-1){6}}
  \put(-140,5){\fcolorbox{black}{white}{\scriptsize{\shortstack[l]{Load cell}}}}
  \put(-120,15){\textcolor{white}{\vector(0,1){20}}}
  \put(-225,5){\fcolorbox{black}{white}{\scriptsize{\shortstack[l]{Switch 1 to\\VNA port 1}}}}
  \put(-200,15){\textcolor{white}{\vector(0,1){42}}}
  \put(-50,5){\fcolorbox{black}{white}{\scriptsize{\shortstack[l]{Switch 2 to\\VNA port 2}}}}
  \put(-30,15){\textcolor{white}{\vector(0,1){42}}}
  \caption{\textbf{Sensor on the load cell platform}, The actuated indenter which can move up-down to exert force on the sesnor, as well as left-right, to do so at a particular location, shown via blue arrows . A load cell below the platform and VNA (not shown here) provide force, phase ground truth measurements.}
  \label{fig:sensor_close_view}
\end{figure}

\begin{table*}[ht]
  \begin{center}
  \begin{tabular}{cccc|c}
    \toprule
     &  
     \hspace{-15mm}
     \begin{minipage}{.1\linewidth}
    \centering
    \includegraphics[width=\linewidth]{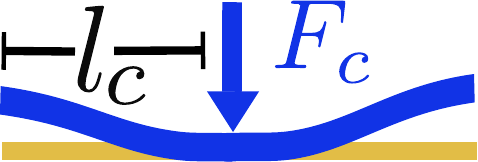}
    \end{minipage}
    
    \hspace{5mm}
     {\small $l_c$ = 20~mm}
     & \hspace{6mm} 
     {\small $l_c$ = 40~mm}& 
     \hspace{9mm} 
     {\small $l_c$=60~mm}& 
     \hspace{-1mm} 
     \hspace{4mm} 
     {\small $l_c$=55~mm (intmd. loc)}\\ \midrule
    \rotatebox[origin=c]{90}{900~MHz}
    &
    \begin{minipage}{.21\linewidth}
    \centering
%
%
\definecolor{mycolor1}{rgb}{1.00000,1.00000,0.00000}%
\begin{tikzpicture}

\begin{axis}[%
width=0.95in,
height=1in,
at={(1.297in,1.093in)},
scale only axis,
xmin=1,
xmax=8,
xlabel={Force (N)},
ymin=20,
ymax=80,
ytick={20,40,60},
ylabel style={at={(axis description cs:-0.13,0.425)},anchor=south},
ylabel={Phase (degrees)},
axis background/.style={fill=white},
xmajorgrids,
ymajorgrids,
legend style={at={(0.41,0.75)}, anchor=south west, legend cell align=left, align=left, draw=white!15!black},
clip mode=individual
]
\addplot [color=cyan, mark=*, mark size=1.5pt, mark options={solid, cyan}]
  table[row sep=crcr]{%
1.899632	71.4624018413735\\
2.195886	69.4205879150598\\
2.497922	74.2007379975595\\
2.79398	73.8909490668953\\
3.142762	72.0528179963408\\
3.52417800000001	64.5734604500948\\
3.91265	63.8607793331772\\
4.814054	65.0478754166988\\
5.381474	63.5211264964515\\
5.918612	65.9228535106582\\
6.45085	62.7783032315911\\
7.002394	64.9013491337717\\
7.526302	63.4704429082359\\
};

\addplot [color=cyan, mark=*, mark size=1.5pt, mark options={solid, cyan}]
  table[row sep=crcr]{%
1.899632	33.1034891514865\\
2.195886	38.6282484061241\\
2.497922	38.7572518734425\\
2.79398	41.4699453367131\\
3.142762	41.750852700428\\
3.524178	40.4465756000586\\
3.91265	41.6844154969915\\
4.814054	42.0113804207021\\
5.381474	48.0231804558347\\
5.918612	48.27541207066\\
6.45085	44.5544358827929\\
7.002394	50.3848254195906\\
7.526302	46.9818370076362\\
};

\addplot [color=brown, mark=*, mark size=1.5pt, mark options={solid, brown}]
  table[row sep=crcr]{%
1.798202	72.665116104409\\
2.07348399999999	68.6241719115095\\
2.36229	68.1743173004841\\
2.711072	72.6909600862476\\
3.055836	65.7712330634573\\
3.41922	65.4698615437667\\
3.88060400000001	67.3785991778349\\
4.256336	67.1102420589379\\
4.707724	65.3946932990578\\
5.15431	66.7094630714085\\
5.722416	65.4316199789454\\
6.244266	65.9025334715852\\
6.748966	65.0711742689263\\
7.296296	63.2808527826019\\
7.734846	65.5492059597359\\
};

\addplot [color=brown, mark=*, mark size=1.5pt, mark options={solid, brown}]
  table[row sep=crcr]{%
1.798202	34.4169185716394\\
2.073484	32.2034770187292\\
2.36229	34.8286955947726\\
2.711072	42.942602921653\\
3.055836	41.0968192801228\\
3.41922	41.0879862426979\\
3.880604	42.2468897371945\\
4.256336	42.2871028490943\\
4.707724	42.2113671656906\\
5.15431	45.5487144753723\\
5.722416	46.2231188106971\\
6.244266	47.650509454328\\
6.748966	50.4841791211098\\
7.296296	45.284368477445\\
7.734846	52.4336620234447\\
};

\addplot [color=black!60!green, mark=*, mark size=1.5pt, mark options={solid, black!60!green}]
  table[row sep=crcr]{%
2.36229	68.1743173004841\\
2.711072	72.6909600862476\\
3.055836	65.7712330634573\\
3.41922	65.4698615437667\\
3.88060400000001	67.3785991778349\\
4.256336	67.1102420589379\\
4.707724	65.3946932990578\\
5.15431	66.7094630714085\\
5.722416	65.4316199789454\\
6.244266	65.9025334715852\\
6.748966	65.0711742689263\\
7.296296	63.2808527826019\\
7.734846	65.5492059597359\\
};

\addplot [color=black!60!green, mark=*, mark size=1.5pt, mark options={solid, black!60!green}]
  table[row sep=crcr]{%
2.36229	34.8286955947726\\
2.711072	42.942602921653\\
3.055836	41.0968192801228\\
3.41922	41.0879862426979\\
3.880604	42.2468897371945\\
4.256336	42.2871028490943\\
4.707724	42.2113671656906\\
5.15431	45.5487144753723\\
5.722416	46.2231188106971\\
6.244266	47.650509454328\\
6.748966	50.4841791211098\\
7.296296	45.284368477445\\
7.734846	52.4336620234447\\
};

\addplot [color=black!20!red, line width=3.0pt]
  table[row sep=crcr]{%
1.89394799999999	64.4326021328748\\
2.06143	64.7282015588884\\
2.224208	64.3134851166884\\
2.404822	64.5092702190336\\
2.599842	64.7028386467375\\
2.805446	64.9332033923237\\
3.011246	65.1241212581789\\
3.211656	65.2170447948588\\
3.41579	65.5189676446074\\
3.630214	65.3637557932314\\
3.85924	65.4715675646795\\
4.10012400000001	65.7722951908614\\
4.335618	65.3858441819391\\
4.582284	65.1583568469546\\
4.819444	64.6641755888253\\
5.03916	64.5183447948441\\
5.21752000000001	64.4583396768846\\
5.41920400000001	64.2088254241905\\
5.666948	63.9498048421383\\
5.89617	63.8213266019119\\
6.090602	63.734651003403\\
6.31953	63.5768556638439\\
6.55208400000001	63.5074413572661\\
6.831874	63.3808774801066\\
7.132734	63.3950940096936\\
7.4431	63.088359625929\\
7.716618	62.8993036416832\\
7.96632200000001	62.6007818223197\\
};

\addplot [color=black!20!red, line width=3.0pt]
  table[row sep=crcr]{%
1.893948	30.2939877699715\\
2.06143	30.9223032266413\\
2.224208	31.9559889841397\\
2.404822	33.0604523944971\\
2.599842	34.5761437908661\\
2.805446	36.8911255336455\\
3.011246	37.7521297177248\\
3.211656	37.5838407284237\\
3.41579	39.2633007529442\\
3.630214	40.1942621675566\\
3.85924	41.9116012702138\\
4.100124	42.3135530153285\\
4.335618	42.9610011750366\\
4.582284	43.7489744296365\\
4.819444	44.1677529448062\\
5.03916	44.5620000071561\\
5.21752	44.8445894904046\\
5.419204	45.1813961419099\\
5.666948	45.487901648195\\
5.89617	45.8650266058091\\
6.090602	46.2712345093153\\
6.31953	46.6190329512716\\
6.552084	46.9786032410852\\
6.831874	47.2114491181594\\
7.132734	47.5196320873157\\
7.4431	47.8467511671587\\
7.716618	48.1483243060401\\
7.966322	48.4784638391886\\
};
\end{axis}
\end{tikzpicture}%
    \end{minipage}
    & 
    \begin{minipage}{.21\linewidth}
    \centering
%
%
\definecolor{mycolor1}{rgb}{1.00000,1.00000,0.00000}%
\begin{tikzpicture}

\begin{axis}[%
width=0.95in,
height=1in,
at={(1.297in,1.093in)},
scale only axis,
xmin=1,
xmax=8,
xlabel={Force (N)},
ymin=-50,
ymax=10,
ylabel style={at={(axis description cs:-0.18,0.425)},anchor=south},
ylabel={Phase (degrees)},
axis background/.style={fill=white},
xmajorgrids,
ymajorgrids,
legend style={at={(0.41,0.75)}, anchor=south west, legend cell align=left, align=left, draw=white!15!black},
clip mode=individual
]
\addplot [color=cyan, mark=*, mark size=1.5pt, mark options={solid, cyan}, forget plot]
  table[row sep=crcr]{%
1.50773	-4.95475124733402\\
1.810942	-2.92018674380645\\
2.151394	-3.27236484187078\\
2.524186	-1.17684592435113\\
2.919028	-2.80904910378135\\
3.309558	-1.19551872823082\\
3.795736	-2.09386421605072\\
4.187736	-1.03832744872737\\
4.671268	-2.04117086183086\\
5.262992	0.459670958682153\\
5.862556	0.570442651825829\\
6.447812	0.210659782669026\\
6.835892	-0.185288429216257\\
7.481516	-1.32137070394174\\
};
\addplot [color=cyan, mark=*, mark size=1.5pt, mark options={solid, cyan}, forget plot]
  table[row sep=crcr]{%
1.50773	-39.3950280880355\\
1.810942	-37.1025928426199\\
2.151394	-37.6450691820621\\
2.524186	-32.9910711259627\\
2.919028	-33.8103550956248\\
3.309558	-33.81932405097\\
3.795736	-34.1495482219422\\
4.187736	-30.7915549490956\\
4.671268	-33.5504551993465\\
5.262992	-29.3970646580844\\
5.862556	-30.3363505727145\\
6.447812	-27.8117393152527\\
6.835892	-29.0921034792634\\
7.481516	-28.377609017005\\
};
\addplot [color=brown, mark=*, mark size=1.5pt, mark options={solid, brown}, forget plot]
  table[row sep=crcr]{%
1.29899	-2.79965645615476\\
1.619646	0.595686532149671\\
1.913352	-1.83263580988373\\
2.224894	-0.850199890784779\\
2.636494	-1.05771634681344\\
3.065048	-2.15823043919423\\
3.507028	-2.44237675230578\\
4.002516	-0.839462188454377\\
4.522406	0.0847885193644515\\
5.027106	-1.48483545697262\\
5.527004	1.45703419915715\\
6.175764	0.844685260820981\\
6.723388	-0.163672262952929\\
7.281498	2.07246228276443\\
7.766794	0.036954847601109\\
};
\addplot [color=brown, mark=*, mark size=1.5pt, mark options={solid, brown}, forget plot]
  table[row sep=crcr]{%
1.29899	-34.9139824358656\\
1.619646	-33.8608201247401\\
1.913352	-36.5205832366013\\
2.224894	-33.4823315589117\\
2.636494	-30.1163983337151\\
3.065048	-33.1122426114896\\
3.507028	-34.228443943025\\
4.002516	-32.9211801117965\\
4.522406	-27.701854555487\\
5.027106	-32.3859581694024\\
5.527004	-29.4680568030587\\
6.175764	-30.8283289125649\\
6.723388	-32.5464242719992\\
7.281498	-29.6956854037805\\
7.766794	-27.857898677441\\
};
\addplot [color=black!60!green, mark=*, mark size=1.5pt, mark options={solid, black!60!green}, forget plot]
  table[row sep=crcr]{%
1.633856	-5.53055521420563\\
1.944222	-2.24052167691731\\
2.27752	-2.20404309431038\\
2.679418	-2.28177532330874\\
3.098956	-1.71755644400605\\
3.542798	-0.888600972666595\\
4.030936	-0.378506448869757\\
4.57562	0.479696007155951\\
5.082868	2.58402730035668\\
5.557482	-0.808132827956697\\
6.225646	2.66127836569216\\
6.837656	1.32805407902852\\
7.220346	2.48688587403777\\
7.933884	2.16465648731612\\
};
\addplot [color=black!60!green, mark=*, mark size=1.5pt, mark options={solid, black!60!green}, forget plot]
  table[row sep=crcr]{%
1.633856	-37.312160094346\\
1.944222	-37.5429121001071\\
2.27752	-35.0190320725409\\
2.679418	-33.154246028916\\
3.098956	-31.054163434275\\
3.542798	-32.1935776202564\\
4.030936	-33.6618498865204\\
4.57562	-29.9997636627111\\
5.082868	-29.5123842230085\\
5.557482	-32.124186729311\\
6.225646	-28.9848714535207\\
6.837656	-27.4110151426051\\
7.220346	-29.4053242034257\\
7.933884	-27.5108636968347\\
};
\addplot [color=black!20!red, line width=3.0pt]
  table[row sep=crcr]{%
0.934626	-9.39805885364069\\
1.087114	-7.90410739548213\\
1.239602	-6.80383578316645\\
1.408358	-5.90533099023946\\
1.58711	-5.08277890574504\\
1.767136	-4.37246204452978\\
1.93256	-3.65434120429222\\
2.098474	-3.09934242057889\\
2.26723	-2.70043610361807\\
2.439808	-2.17325860984877\\
2.65139	-1.93740067743812\\
2.881494	-1.59578160734179\\
3.316222	-0.998357223871054\\
3.526726	-0.73887685875107\\
3.965178	-0.237326369051488\\
4.198712	-0.0942102267948712\\
4.415488	0.0911441951729302\\
4.658136	0.30522985826498\\
4.908134	0.531982998124263\\
5.151664	0.711519676499648\\
5.414696	0.828147613467195\\
5.704776	0.9815821289681\\
6.011908	1.11803209036559\\
6.359318	1.27990075573318\\
6.682718	1.45577752821225\\
7.03101	1.60307072757709\\
7.36862	1.78319360939511\\
7.694274	1.86956966229536\\
};
\addplot [color=black!20!red, line width=3.0pt]
  table[row sep=crcr]{%
0.934626000000002	-43.0889319700806\\
1.087114	-41.7633219046514\\
1.239602	-40.4740770941017\\
1.408358	-39.3123216005029\\
1.58711	-38.2854547830293\\
1.767136	-37.2795589100205\\
1.93256	-35.7327250348542\\
2.098474	-34.7822556498596\\
2.26723	-34.1967972856428\\
2.439808	-33.7680866661522\\
2.65139	-33.2741350739774\\
2.881494	-32.8798740960742\\
3.100426	-32.6040590444666\\
3.316222	-32.4224000983938\\
3.526726	-32.1998021982398\\
3.746246	-31.8623510287267\\
3.965178	-31.4761471877231\\
4.198712	-31.295923080691\\
4.415488	-31.2616038453197\\
4.658136	-31.1939551528576\\
4.908134	-31.0493149679663\\
5.151664	-30.9883428781048\\
5.414696	-30.7931851473691\\
5.704776	-30.6845081564844\\
6.011908	-30.4579521060285\\
6.359318	-30.2992469886208\\
6.682718	-30.1877397350036\\
7.03101	-30.0222528917941\\
7.36862	-29.8794159321712\\
7.694274	-29.7653694495184\\
};
\end{axis}
\end{tikzpicture}%
    \end{minipage}
    & 
    \begin{minipage}{.21\linewidth}
    \centering
%
%
\definecolor{mycolor1}{rgb}{1.00000,1.00000,0.00000}%
\begin{tikzpicture}

\begin{axis}[%
width=0.95in,
height=1in,
at={(1.297in,1.093in)},
scale only axis,
xmin=1,
xmax=8,
xlabel={Force (N)},
ymin=-150,
ymax=-90,
ylabel style={at={(axis description cs:-0.25,0.425)},anchor=south},
ylabel={Phase (degrees)},
axis background/.style={fill=white},
xmajorgrids,
ymajorgrids,
legend style={at={(0.41,0.75)}, anchor=south west, legend cell align=left, align=left, draw=white!15!black},
clip mode=individual
]
\addplot [color=cyan, mark=*, mark size=1.5pt, mark options={solid, cyan}, forget plot]
  table[row sep=crcr]{%
2.101806	-113.191916596258\\
2.36983600000001	-111.102910639765\\
2.666678	-106.629649789103\\
2.98135600000001	-109.811390104757\\
3.36012599999999	-108.118750796973\\
3.768198	-108.693650434794\\
4.125506	-106.244822880773\\
4.371878	-107.229015280873\\
4.749668	-105.941387760828\\
5.189982	-105.677343664653\\
5.516812	-105.100461935354\\
5.8653	-103.925556425923\\
6.23966	-104.492163793085\\
6.55874799999999	-102.985884300027\\
6.954472	-103.132632210824\\
7.63959	-102.756803918915\\
};
\addplot [color=cyan, mark=*, mark size=1.5pt, mark options={solid, cyan}, forget plot]
  table[row sep=crcr]{%
2.10180600000001	-145.984938818011\\
2.36983599999999	-144.208740586352\\
2.66667799999999	-140.94007859016\\
2.98135600000001	-139.153129489151\\
3.36012600000001	-146.740687123553\\
3.76819800000001	-145.593793322759\\
4.125506	-143.911375184165\\
4.37187800000001	-142.307957654085\\
4.74966800000001	-142.527906897407\\
5.18998200000001	-142.371162012406\\
5.51681199999999	-142.939448564335\\
5.86529999999999	-141.603765938097\\
6.23966000000001	-142.968676706985\\
6.55874800000001	-142.903814666521\\
6.95447200000001	-142.994723364124\\
7.63959	-145.585528480038\\
};
\addplot [color=black!60!green, mark=*, mark size=1.5pt, mark options={solid, black!60!green}, forget plot]
  table[row sep=crcr]{%
1.861118	-116.720547233253\\
2.153746	-115.521836389706\\
2.456664	-113.300870230946\\
2.74292199999999	-111.107078251714\\
3.024672	-108.953314269012\\
3.366496	-108.836938225128\\
3.72468600000001	-108.560039254709\\
4.06504	-106.681078028499\\
4.37521	-106.025858159097\\
4.712036	-105.278321952231\\
5.006428	-106.843023774242\\
5.41597	-104.059804878955\\
5.831784	-104.335779647006\\
6.129116	-104.934404127123\\
6.68938199999999	-102.766760850211\\
7.10372599999999	-101.471534092583\\
7.678104	-102.509222792677\\
};
\addplot [color=black!60!green, mark=*, mark size=1.5pt, mark options={solid, black!60!green}, forget plot]
  table[row sep=crcr]{%
1.861118	-142.857046691678\\
2.15374600000001	-143.572241803628\\
2.45666399999999	-143.384565375382\\
2.74292199999999	-147.109633221407\\
3.02467200000001	-145.116808002536\\
3.36649600000001	-144.376231584361\\
3.72468599999999	-141.275376891306\\
4.06504000000001	-144.872921940021\\
4.37521000000001	-145.142945216782\\
4.71203600000001	-143.953278337892\\
5.006428	-146.999429416621\\
5.41596999999999	-139.651706972176\\
5.831784	-142.679534739686\\
6.12911600000001	-142.856529175452\\
6.68938199999999	-142.283560965452\\
7.10372599999999	-146.919536547046\\
7.67810399999999	-137.643609331856\\
};
\addplot [color=brown, mark=*, mark size=1.5pt, mark options={solid, brown}, forget plot]
  table[row sep=crcr]{%
1.896986	-114.549029494791\\
2.215976	-114.682582509539\\
2.49115999999999	-112.323828900996\\
2.777418	-111.450607534721\\
3.021536	-111.46178084752\\
3.33004	-108.572873198895\\
3.61306399999999	-109.009745232437\\
3.871196	-106.504434638623\\
4.24937800000001	-106.435466226685\\
4.61972	-104.745102774856\\
5.030732	-104.589570261821\\
5.44664400000001	-105.305163372925\\
5.87862800000001	-105.190248448451\\
6.22378399999999	-101.762480328862\\
6.69928	-101.842364792119\\
7.221718	-100.636402002932\\
7.574224	-101.027043059838\\
};
\addplot [color=brown, mark=*, mark size=1.5pt, mark options={solid, brown}, forget plot]
  table[row sep=crcr]{%
1.896986	-146.036290796676\\
2.21597600000001	-146.543388924889\\
2.49116000000001	-142.420996310153\\
2.77741800000001	-142.423963533065\\
3.021536	-144.174172186275\\
3.33004	-140.870003663836\\
3.61306400000001	-143.16158379146\\
3.871196	-143.868299539317\\
4.24937800000001	-143.416458425074\\
4.61972	-144.144766213778\\
5.030732	-143.290272874168\\
5.44664399999999	-143.651868191812\\
5.87862799999999	-142.117021140348\\
6.22378399999999	-140.931341787736\\
6.69927999999999	-140.398243356265\\
7.22171800000001	-143.369098325602\\
7.57422399999999	-144.387468003474\\
};
\addplot [color=black!20!red, line width=3.0pt]
  table[row sep=crcr]{%
1.917762	-117.793079859696\\
2.077698	-116.353795781266\\
2.217838	-115.150173113253\\
2.35935000000001	-114.205539683021\\
2.494688	-113.241764305137\\
2.64061	-112.447616728168\\
2.79839	-111.646693113387\\
2.974692	-110.846939285751\\
3.12963000000001	-110.077598968945\\
3.28574399999999	-109.478657378499\\
3.43882000000001	-108.825405876819\\
3.603264	-108.334176837321\\
3.76329800000001	-107.762917553481\\
3.925586	-107.267469178838\\
4.121978	-106.800053820582\\
4.327484	-106.299515481922\\
4.50163000000001	-105.773314679149\\
4.684988	-105.320517460653\\
4.858056	-104.877025327372\\
5.03769	-104.457560019007\\
5.173028	-104.047629966184\\
5.278868	-103.715066771132\\
5.490156	-103.508834854808\\
5.79709200000001	-103.153582967515\\
6.03827	-102.667220500804\\
6.247794	-102.301943193531\\
6.469078	-101.997615264218\\
6.70702199999999	-101.710552389828\\
6.97123000000001	-101.343777207549\\
7.24200399999999	-101.043637795199\\
7.501508	-100.751268342665\\
7.74494	-100.489533493325\\
};
\addplot [color=black!20!red, line width=3.0pt]
  table[row sep=crcr]{%
1.91776200000001	-143.555819485566\\
2.077698	-143.262992020963\\
2.217838	-142.943678716936\\
2.35935000000001	-142.702511607477\\
2.494688	-142.520732860777\\
2.79839000000001	-142.291393178475\\
2.974692	-142.235506051506\\
3.12962999999999	-142.171432144573\\
3.28574399999999	-142.160036082988\\
3.43881999999999	-142.101792713732\\
3.603264	-142.165020843888\\
3.76329799999999	-142.189780438986\\
3.92558600000001	-142.180083820631\\
4.12197800000001	-142.201371900557\\
4.327484	-142.268880725606\\
4.50163000000001	-142.243161418288\\
4.684988	-142.298596717381\\
4.858056	-142.355680633337\\
5.03769	-142.330758775938\\
5.17302799999999	-142.408026713959\\
5.27886799999999	-142.49203662468\\
5.49015600000001	-142.516293245437\\
5.79709199999999	-142.601034134681\\
6.03827000000001	-142.613892373368\\
6.247794	-142.686746729625\\
6.70702199999999	-142.857726964741\\
6.97122999999999	-142.842874215342\\
7.24200400000001	-142.931619744362\\
7.501508	-143.030285481531\\
7.74494000000001	-143.14629880225\\
};
\end{axis}

\end{tikzpicture}%
    \end{minipage}
    & 
    \begin{minipage}{.21\linewidth}
    \centering
    \input{results/all_wireless_900_55_fin.tex}
    \end{minipage}
    \\ \midrule
    \rotatebox[origin=c]{90}{2.4~GHz}
    &
    \begin{minipage}{.21\linewidth}
    \centering
%
%
\definecolor{mycolor1}{rgb}{1.00000,1.00000,0.00000}%
\begin{tikzpicture}

\begin{axis}[%
width=0.95in,
height=1in,
at={(1.297in,1.093in)},
scale only axis,
xmin=1,
xmax=8,
xlabel={Force (N)},
ymin=10,
ymax=120,
ytick={20,40,60,80,100},
ylabel style={at={(axis description cs:-0.15,0.425)},anchor=south},
ylabel={Phase (degrees)},
axis background/.style={fill=white},
xmajorgrids,
ymajorgrids,
legend style={at={(0.421,0.744)}, anchor=south west, legend cell align=left, align=left, draw=white!15!black},
clip mode=individual
]
\addplot [color=cyan, mark=*, mark size=1.5pt, mark options={solid, cyan}]
  table[row sep=crcr]{%
2.371306	113.204006359661\\
2.485868	112.924281407025\\
2.630026	112.227249040395\\
2.77820199999999	109.299208028379\\
2.948134	113.95143921656\\
3.166478	109.098498226472\\
3.39138800000001	114.477092184942\\
3.563672	100.095545641142\\
3.79759799999999	102.503377861155\\
3.943324	97.6908598585457\\
4.073076	95.4649562871288\\
4.30562999999999	100.087570623617\\
4.589732	100.230751081482\\
4.82552	102.586641343003\\
4.981144	104.485483181059\\
5.078262	103.390504479081\\
5.27142000000001	102.295525777102\\
5.51495	101.900388665715\\
5.615008	98.5116372727611\\
5.937918	96.9232683049799\\
6.09746200000001	98.3698832900438\\
6.30042	99.6476899369839\\
6.60284799999999	98.4891814247223\\
6.795124	98.6851871468993\\
7.058646	99.763645058718\\
7.519246	95.3554275446876\\
7.651252	99.8052709023669\\
7.83539399999999	100.094958134706\\
8.05364	103.700659021019\\
};

\addplot [color=cyan, mark=*, mark size=1.5pt, mark options={solid, cyan}]
  table[row sep=crcr]{%
2.371306	15.4193997537209\\
2.485868	15.2279546274187\\
2.630026	20.0712801253462\\
2.778202	21.013002407121\\
2.948134	23.0646483815678\\
3.166478	30.5354471324753\\
3.391388	37.1979472160529\\
3.563672	29.0727343894761\\
3.797598	35.8677246222693\\
3.943324	40.0991790515704\\
4.073076	43.0786481336007\\
4.30563	38.7148450448295\\
4.589732	45.9145543109286\\
4.82552	48.753932524156\\
4.981144	43.1936926795607\\
5.078262	47.1470278499687\\
5.27142	48.2002789708574\\
5.51495	46.7216263599416\\
5.615008	52.7087553270358\\
5.937918	54.5817657853495\\
6.097462	53.586890231215\\
6.30042	51.486390527008\\
6.602848	53.1918531436656\\
6.795124	58.7666084918398\\
7.058646	56.7250119667718\\
7.519246	61.3881586058973\\
7.651252	60.038633408228\\
7.835394	57.2525366725755\\
8.05364	63.688782506204\\
};

\addplot [color=brown, mark=*, mark size=1.5pt, mark options={solid, brown}]
  table[row sep=crcr]{%
2.565248	109.415924111317\\
2.729496	111.942727862866\\
2.922752	111.997115748148\\
3.074652	106.003016032547\\
3.30162	112.63949789249\\
3.488212	92.5804845724299\\
3.670198	104.875542514182\\
3.92656599999999	100.691311381989\\
4.112178	98.3883553966483\\
4.33601	107.316842642531\\
4.49575	103.304320709725\\
4.70664600000001	103.878709155435\\
4.92744	103.984928299988\\
5.19772399999999	99.9920681426124\\
5.504366	102.313796060025\\
5.722318	104.635523977438\\
5.99926600000001	99.3364391702698\\
6.24220800000001	100.471638908778\\
6.505926	103.9127184254\\
6.77287800000001	100.072176127655\\
7.063056	100.83587340839\\
7.34216000000001	95.3695724614485\\
7.579222	96.8079453743541\\
7.85401400000001	95.9426428862855\\
8.074906	94.6069587093021\\
};

\addplot [color=brown, mark=*, mark size=1.5pt, mark options={solid, brown}]
  table[row sep=crcr]{%
2.565248	18.978717161699\\
2.729496	20.709494071164\\
2.922752	21.6351815707667\\
3.074652	27.1952631373662\\
3.30162	33.0561921740833\\
3.488212	28.1280494052407\\
3.670198	34.3674096833432\\
3.926566	38.7949389215798\\
4.112178	40.3858864384477\\
4.33601	43.9087549671253\\
4.49575	47.141093779029\\
4.706646	46.629880521143\\
4.92744	50.8313504100793\\
5.197724	49.2650027431551\\
5.504366	49.9490045691827\\
5.722318	52.4198960840274\\
5.999266	52.4869487664133\\
6.242208	54.4010454997513\\
6.505926	56.2121406697104\\
6.772878	58.0063922389681\\
7.063056	57.6978592404317\\
7.34216	53.5117885030767\\
7.579222	61.0993794712383\\
7.854014	59.2898234929047\\
8.074906	60.3284337244404\\
};

\addplot [color=black!60!green, mark=*, mark size=1.5pt, mark options={solid, black!60!green}]
  table[row sep=crcr]{%
3.03506	111.793785520612\\
3.21881	107.333798501231\\
3.41334000000001	93.9238599530803\\
3.59189600000001	93.5891879347445\\
3.76114200000001	98.1318902922956\\
3.962238	95.0841731294885\\
4.16931200000001	98.6123861592755\\
4.36639	101.192532156441\\
4.61678000000001	100.186288523549\\
4.76407399999999	100.450027141911\\
4.982124	104.673115645988\\
5.324438	102.826889775294\\
5.600798	99.9979377480832\\
5.704384	98.7538241581933\\
6.02092399999999	97.5097105683033\\
6.33031	103.62618927578\\
6.571684	98.4737105990925\\
6.801984	96.0305124828014\\
7.023268	92.3882306636939\\
7.37126600000001	100.407592664638\\
7.53600400000001	94.0144160149602\\
7.997192	98.0602114770585\\
};

\addplot [color=black!60!green, mark=*, mark size=1.5pt, mark options={solid, black!60!green}]
  table[row sep=crcr]{%
3.03506	29.2373029766891\\
3.21881	26.7577644489429\\
3.41334	37.586070863613\\
3.591896	33.4948820256047\\
3.761142	35.91874582\\
3.962238	41.2602083732278\\
4.169312	41.598571206422\\
4.36639	43.5696174037598\\
4.61678	36.9799220046087\\
4.764074	44.4095898094567\\
4.982124	46.3715500714923\\
5.324438	53.1079429142118\\
5.600798	53.3729946218813\\
5.704384	47.9198795235441\\
6.020924	54.2222651740463\\
6.33031	57.3437383052691\\
6.571684	54.4971874791499\\
6.801984	56.9754369709215\\
7.023268	56.6064265294841\\
7.371266	62.8107807837249\\
7.536004	56.539768414371\\
7.997192	63.6795115532638\\
};

\addplot [color=black!20!red, line width=3.0pt]
  table[row sep=crcr]{%
3.25859800000001	100.604578409545\\
3.346896	99.3683451473795\\
3.444014	103.669234943543\\
3.535938	101.661432965764\\
3.627568	101.713009267002\\
3.71723799999999	104.491028741769\\
3.813768	100.152069376838\\
3.90559399999999	101.60685092147\\
4.00075200000001	101.858200089993\\
4.100712	100.938336844851\\
4.22282	101.192060606286\\
4.334344	99.1746205854268\\
4.443516	98.2278918853767\\
4.54857200000001	97.267722759122\\
4.662644	98.6054449729527\\
4.775246	99.0250663846253\\
4.883928	98.5340054998351\\
4.994668	99.2346144134051\\
5.101292	99.3906191094028\\
5.212718	99.4894083914359\\
5.32502600000001	99.4859866519054\\
5.44546800000001	99.7557159561982\\
5.559344	100.233482194361\\
5.679982	100.422789721867\\
5.796994	101.109059713899\\
5.920376	101.439549758502\\
6.04219000000001	101.576281635269\\
6.164886	101.598219984798\\
6.290718	101.118512254256\\
6.417334	101.054012660209\\
6.531994	100.915137495336\\
6.65508200000001	101.099876494113\\
6.775524	101.23541234759\\
6.892928	101.103783275436\\
7.00503999999999	101.138679241915\\
7.117642	101.274746042011\\
7.22103200000001	101.5991727258\\
7.333536	101.580722715473\\
7.45231200000001	101.483964875741\\
7.57412600000001	101.377356787876\\
7.687022	101.24640173907\\
7.80423	101.305112111997\\
7.92281	101.408052917471\\
};

\addplot [color=black!20!red, line width=3.0pt]
  table[row sep=crcr]{%
3.258598	30.5209084707355\\
3.346896	32.334906943887\\
3.444014	32.4741609215864\\
3.535938	33.777946626305\\
3.627568	35.4354295682238\\
3.717238	35.7612590917641\\
3.813768	36.4903885513975\\
3.905594	37.3868967776766\\
4.000752	37.7444993420038\\
4.100712	38.2195168510523\\
4.22282	39.4260045859801\\
4.334344	40.0704416465426\\
4.443516	41.2368739695343\\
4.548572	42.0132467831974\\
4.662644	42.9466772886497\\
4.775246	43.6781921696305\\
4.883928	44.2611134702947\\
4.994668	45.0585780777911\\
5.101292	45.9616469750679\\
5.212718	46.7044141899093\\
5.325026	48.0957739450127\\
5.445468	48.4710220339729\\
5.559344	48.9965877774389\\
5.679982	50.8983341768474\\
5.796994	52.2763252942718\\
5.920376	52.8214798924534\\
6.04219	53.2857324046777\\
6.164886	52.5565702262962\\
6.290718	55.3470409079548\\
6.417334	55.3340737377652\\
6.531994	56.1868985894092\\
6.655082	56.8445762620494\\
6.775524	57.4509024558687\\
6.892928	57.895717612878\\
7.00504	58.2498752377616\\
7.117642	58.6795370619022\\
7.221032	59.2350554931772\\
7.333536	59.8240063542901\\
7.452312	59.9668562464531\\
7.574126	60.3685837215411\\
7.687022	60.6475760020533\\
7.80423	61.1442531510639\\
7.92281	61.5245187496275\\
};

\end{axis}
\end{tikzpicture}%
    \end{minipage}
    & 
    \begin{minipage}{.21\linewidth}
    \centering
%
%
\definecolor{mycolor1}{rgb}{1.00000,1.00000,0.00000}%
\begin{tikzpicture}

\begin{axis}[%
width=0.95in,
height=1in,
at={(1.297in,1.093in)},
scale only axis,
xmin=1,
xmax=8,
xlabel={Force (N)},
ymin=-120,
ymax=-10,
ytick={-120,-100,-80,-60,-40,-20},
ylabel style={at={(axis description cs:-0.23,0.425)},anchor=south},
ylabel={Phase (degrees)},
axis background/.style={fill=white},
xmajorgrids,
ymajorgrids,
legend style={at={(0.421,0.744)}, anchor=south west, legend cell align=left, align=left, draw=white!15!black},
clip mode=individual
]
\addplot [color=cyan, mark=*, mark size=1.5pt, mark options={solid, cyan}, forget plot]
  table[row sep=crcr]{%
1.28037	-52.8630875550957\\
1.842498	-51.7270084615495\\
2.018702	-45.9657863928037\\
2.237634	-46.4143630855143\\
2.507624	-44.5584018747463\\
2.750958	-40.514593721635\\
2.98263	-42.7649164323333\\
3.226454	-41.7104946438573\\
3.509086	-40.4317210610026\\
3.820432	-35.8380936519067\\
4.06798	-39.5611688042043\\
4.360118	-30.5203510818995\\
4.711448	-28.0258305146255\\
5.02691	-35.5062494423312\\
5.34051	-29.7170942290012\\
5.620104	-27.5106827298955\\
6.017788	-32.2018037548809\\
6.341482	-27.4148317186458\\
6.633914	-20.9079558545539\\
7.024836	-15.8597051250237\\
7.32403	-18.8972778402218\\
7.71897	-21.9348505554199\\
8.075396	-22.0382791734136\\
};
\addplot [color=cyan, mark=*, mark size=1.5pt, mark options={solid, cyan}, forget plot]
  table[row sep=crcr]{%
1.28037	-111.180505893154\\
1.84249800000001	-105.721266664752\\
2.018702	-108.18710016975\\
2.237634	-107.337877569717\\
2.50762400000001	-103.214935757773\\
2.750958	-103.097911455712\\
2.98263	-102.94647177989\\
3.226454	-99.6016757271263\\
3.509086	-97.3047180848286\\
3.820432	-90.6556305338697\\
4.06798000000001	-91.787344299156\\
4.360118	-88.9843120649765\\
4.711448	-88.6740812489749\\
5.02691	-84.6747977964213\\
5.34050999999999	-86.0110171265285\\
5.620104	-87.2630230084088\\
6.017788	-90.8802478762581\\
6.341482	-85.8173493388984\\
6.633914	-84.4453373776489\\
7.02483600000001	-84.4293694731822\\
7.32402999999999	-82.1796396219344\\
7.71897	-79.9299097706865\\
8.075396	-82.9721857817805\\
};
\addplot [color=brown, mark=*, mark size=1.5pt, mark options={solid, brown}, forget plot]
  table[row sep=crcr]{%
1.591422	-43.9211153803684\\
2.248316	-47.0279804027556\\
2.460192	-46.8634523970973\\
2.72391	-50.6475441218831\\
2.981356	-44.9750417127661\\
3.23302	-41.2769589372611\\
3.496248	-45.5690133613603\\
3.795442	-40.6232244815092\\
4.103848	-37.2698514494178\\
4.373642	-32.7311624805812\\
4.729872	-28.9127229232307\\
5.082868	-32.3591169056427\\
5.443802	-34.2629007908545\\
5.732706	-34.6238086903938\\
6.129802	-22.3511302977448\\
6.492696	-21.0085206128908\\
6.840498	-19.6235779614683\\
7.133126	-17.1172009073107\\
7.532476	-18.7032200326072\\
7.839804	-19.5185170350192\\
8.218672	-20.3338140374313\\
};
\addplot [color=brown, mark=*, mark size=1.5pt, mark options={solid, brown}, forget plot]
  table[row sep=crcr]{%
1.59142199999999	-102.306340128872\\
2.248316	-106.784916665748\\
2.46019200000001	-103.806934057658\\
2.72391	-105.080622607561\\
2.98135600000001	-97.9687718175551\\
3.23302	-96.6448968303266\\
3.49624799999999	-93.0215248298755\\
3.79544199999999	-89.6858306896911\\
4.103848	-92.1071710398893\\
4.373642	-86.6593742266337\\
4.729872	-91.3848021971669\\
5.082868	-87.5771842880051\\
5.44380200000001	-85.5991319210158\\
5.73270600000001	-90.7492204609254\\
6.129802	-82.2484388487872\\
6.492696	-85.2485504472964\\
6.840498	-84.2176009429363\\
7.133126	-81.450184534086\\
7.532476	-81.60922873489\\
7.839804	-83.1900156472435\\
8.218672	-84.7708025595971\\
};
\addplot [color=black!60!green, mark=*, mark size=1.5pt, mark options={solid, black!60!green}, forget plot]
  table[row sep=crcr]{%
1.3426	-51.2111854297532\\
1.962646	-51.2663111202665\\
2.157274	-47.8977163502823\\
2.397472	-48.0716362463832\\
2.649528	-48.8212600971522\\
2.914422	-38.2523013193505\\
3.175004	-38.8886046909852\\
3.463124	-36.5456506572564\\
3.730272	-41.6403158137781\\
4.01849	-35.078485541858\\
4.31543	-37.2596672202071\\
4.610116	-34.8672281950901\\
4.96762	-36.9412680284132\\
5.338256	-28.5895566766309\\
5.68155	-33.3529352017924\\
6.018376	-27.2074282537032\\
6.410376	-19.6279663127604\\
6.801396	-14.4346993032879\\
7.154196	-19.9201598293477\\
7.516894	-27.8862545123652\\
7.84588	-23.5479766497043\\
8.124102	-20.9615148114408\\
};
\addplot [color=black!60!green, mark=*, mark size=1.5pt, mark options={solid, black!60!green}, forget plot]
  table[row sep=crcr]{%
1.3426	-108.026825160289\\
1.96264600000001	-111.306015860166\\
2.157274	-105.805444461396\\
2.39747199999999	-98.4980702276756\\
2.649528	-104.160349479712\\
2.914422	-104.714150471655\\
3.175004	-106.209928250181\\
3.46312399999999	-100.417244501765\\
3.730272	-92.7177709109952\\
4.01849	-92.8942366727339\\
4.31543000000001	-92.2505187571575\\
4.610116	-86.8708384309795\\
4.96762	-89.7562075698658\\
5.338256	-87.4824591857064\\
5.68155	-85.2205641597401\\
6.018376	-86.2453408369563\\
6.410376	-80.3184395654005\\
6.801396	-77.5174861660245\\
7.154196	-77.0646755689317\\
7.51689399999999	-78.4584108252957\\
7.84587999999999	-79.8558436650282\\
8.12410200000001	-80.7617559848281\\
};
\addplot [color=black!20!red, line width=3.0pt]
  table[row sep=crcr]{%
1.290954	-56.7099460700665\\
1.366218	-52.7062433144049\\
1.464904	-49.4112485887025\\
1.558004	-48.4878202244818\\
1.666882	-50.5718097515347\\
1.764784	-50.5988558816327\\
1.861608	-49.3134750783011\\
1.964508	-49.2977498991289\\
2.206176	-45.9229993126809\\
2.319758	-44.654744255713\\
2.426774	-45.2544408081703\\
2.526832	-43.0728906090006\\
2.635416	-41.0058227872327\\
2.733024	-40.0253976361076\\
2.835924	-39.6210347756455\\
2.954308	-41.7907136938612\\
3.068478	-39.6912659327326\\
3.19823	-40.7286607909673\\
3.319848	-36.8560388132644\\
3.44323	-36.2855729312335\\
3.575236	-36.1421792183812\\
3.692248	-35.1584048189022\\
3.807594	-34.443914130205\\
3.924508	-34.3121637334821\\
4.045342	-35.6613579772744\\
4.157258	-34.1504756867819\\
4.282404	-33.7148506127452\\
4.40412	-33.3917278249785\\
4.522798	-32.9391040155459\\
4.630402	-32.4526342445595\\
4.764858	-32.2105687126173\\
4.882556	-31.9823860783776\\
5.018482	-31.7021703036188\\
5.140982	-30.2977028128744\\
5.25819	-29.4253832081075\\
5.372752	-29.4027661269024\\
5.497898	-28.2465402114009\\
5.610794	-27.7070332503518\\
5.731922	-27.7092727617492\\
5.851188	-26.3735685624684\\
5.970552	-26.8760642130527\\
6.079136	-26.7012199917813\\
6.196148	-26.4115336109327\\
6.303948	-26.0039028975359\\
6.41704	-25.8034412394933\\
6.527976	-25.2976646796937\\
6.650574	-25.0246774244586\\
6.764254	-24.6420196092016\\
6.879796	-24.4497453929237\\
6.99671	-24.1399108481431\\
7.114408	-23.6281591765567\\
7.21623	-23.103748270678\\
7.328342	-22.6479992241304\\
7.433202	-22.3227185944267\\
7.55384	-22.2468664198536\\
7.649292	-22.0309048675383\\
7.77091	-21.9720024389427\\
7.870184	-21.7616159697598\\
7.991214	-21.6843101373534\\
};
\addplot [color=black!20!red, line width=3.0pt]
  table[row sep=crcr]{%
1.290954	-110.592598998966\\
1.366218	-106.447706640407\\
1.464904	-106.073712372379\\
1.558004	-108.768757761506\\
1.666882	-109.447560491208\\
1.76478400000001	-105.98950922432\\
1.861608	-104.756681277174\\
1.964508	-103.561259271156\\
2.085244	-103.15107936919\\
2.206176	-102.953165373072\\
2.31975800000001	-103.254608824624\\
2.42677399999999	-102.199990584037\\
2.526832	-101.953641162766\\
2.63541600000001	-101.256574054942\\
2.733024	-99.504448226429\\
2.83592400000001	-98.3405543152656\\
2.954308	-97.9991838572391\\
3.068478	-96.3136806380318\\
3.19823	-96.3044464230728\\
3.31984799999999	-96.3869452895946\\
3.44323	-96.136430373575\\
3.575236	-94.8663479933044\\
3.69224800000001	-94.063094971392\\
3.80759399999999	-93.2066703231569\\
3.924508	-93.1067575785978\\
4.04534200000001	-93.5393817554407\\
4.157258	-94.2716100190906\\
4.282404	-93.6523526541774\\
4.40412000000001	-93.2934737739303\\
4.52279799999999	-92.7171774364425\\
4.630402	-91.4302630253211\\
4.764858	-90.1071368011459\\
4.88255599999999	-88.8720159271311\\
5.01848200000001	-89.2825716259142\\
5.14098200000001	-89.598274807438\\
5.25819	-89.3481095084057\\
5.37275200000001	-88.8063396002774\\
5.49789800000001	-88.4355362959201\\
5.610794	-88.0288220577552\\
5.731922	-86.0100353213744\\
5.85118800000001	-86.3660440106774\\
5.970552	-86.4693078011363\\
6.07913600000001	-85.7877206986338\\
6.19614799999999	-84.8331026565263\\
6.30394800000001	-84.901297269074\\
6.41704	-84.4325427168968\\
6.527976	-83.9862082436683\\
6.65057400000001	-83.8929461436556\\
6.76425400000001	-82.0561754794927\\
6.879796	-80.6051092136589\\
6.99671000000001	-80.6578542220018\\
7.114408	-80.4681690173769\\
7.21623	-79.6521348153358\\
7.32834200000001	-79.3825524114008\\
7.43320199999999	-79.4997390758912\\
7.55383999999999	-79.4390914262013\\
7.649292	-79.1970991139705\\
7.77091	-79.2119648858022\\
7.87018399999999	-79.1016166046407\\
7.991214	-78.9378474846534\\
};
\end{axis}
\end{tikzpicture}%
    \end{minipage}
    &
    \begin{minipage}{.21\linewidth}
    \centering
    \input{results/all_wireless_2400_60_fin.tex}
    \end{minipage}
    & 
    \begin{minipage}{.21\linewidth}
    \centering
    \input{results/all_wireless_2400_55_fin.tex}
    \end{minipage}\\
    \midrule
    \multicolumn{5}{l}{\small Legend \hspace{4mm} 
    \hspace{4mm}
    \textcolor{red}{\textbf{---}} 
    {\scriptsize VNA Phases} \hspace{8mm} \tikz\draw[cyan,fill=cyan] (0,0) circle (.75ex); {\scriptsize Wireless trial 1} \hspace{8mm} \tikz\draw[brown,fill=brown] (0,0) circle (.75ex); 
    {\scriptsize Wireless trial 2} \hspace{8mm} \tikz\draw[black!60!green,fill=black!60!green] (0,0) circle (.75ex); 
    {\scriptsize Wireless trial 3} \hspace{8mm} \textcolor{magenta}{\textbf{---}} 
    {\scriptsize Model}} \\
    \bottomrule
  \end{tabular}
  \caption{Ground truth phase-force profiles (red) measured via the VNA and load cell setup show symmetric phase changes when pressed at center ($l_c=40$mm, total length is 80mm), and asymmetric phases when pressed at $l_c=20,60$mm, as discussed in Section~\ref{subsec:force_transduction}. We collect VNA data at $l_c=20, 30, 40, 50, 60$mm, and perform a cubic fit to get the sensor model, which is evaluated by testing at intmd. $l_c=55$mm.
  The wireless phase measurements, as well as the model predictions at $55$mm consistently overlap with ground truth, warranting the performance of \name's design .\vspace{-3mm}}
\label{table:open_closed_loop_figures}
\end{center}
\end{table*}

To verify if the sensor is following the transduction mechanism, we exert forces on the sensor at 5 locations, namely 20, 30, 40, 50 and 60~mm (20, 40, 60 marked in Fig.~\ref{fig:sensor_close_view}). We expect the beam to show a symmetric phase changes on both the ends when tested at the center point, and asymmetric phase changes for the end points, as described in Section~\ref{subsec:force_transduction}. When pressed on the end points, the port near the pressing location would show more phase change, whereas the other end essentially shows a constant phase as force increases. For this testing, we use the setup visible Fig.~\ref{fig:sensor_close_view}, where an indenter allows us to apply a force at a given location on the sensor, and a load cell on which the sensor is attached, allows us to collect the values of force magnitudes applied. As seen clearly in the 20/40/60~mm figures in Table~\ref{table:open_closed_loop_figures}, the phases do follow the beam bending model as discussed above, since 40~mm testing shows symmetric behaviour, whereas for 20/40~mm, one of the ends show a constant phase as force magnitude is increased. 

We now use the data obtained by applying forces at all 5 locations, and compute a cubic-fit to make a sensor model that allows us to compute the force magnitude and force location based on the measured phase changes. To confirm the validity of the model, we asses it at an intermediate point (55~mm), and plot the phase-force profile as predicted by the model alongside the ground-truth profile we collect from the VNA. As visible in Table~\ref{table:open_closed_loop_figures}, all graphs for force applied at 55~mm overlay on each other, which confirms the reliability of the sensor model.

\subsection{Clock Design and RF Switches}

\noindent To encode the phase changes caused by different shorting positions on the microstrip line due to application of a force, we utilize $2$ RF-switches with the duty cycled clocking strategy described in Section~\ref{Subsec:double-ended}. 
We use the HMC544AE from Analog Devices in our prototype, which is a reflective-open switch consistent with our duty cycling requirements discussed in Section~\ref{Subsec:double-ended}.

The final component in our prototype design is the clock source. We use the timer channels in Arduino Due with an Atmel SAM3X8E ARM Cortex-M3 processor\cite{ardue} to generate the duty cycled clock source as described in Section 2. We generate a $25\%$ duty cycled $1$~kHz square wave, and a $75\%$ duty cycled $2$~kHz square wave to modulate the two RF switches. This gives us interference-free modulation at $1,4$~kHz. 

Hence our sensor prototype consists of five components, shown in Fig.~\ref{fig:gelatin_experiment}, the microstrip line sensor, 2 RF switches, 2 clock sources, a splitter to combine outputs of the 2 switches and one antenna to communicate the backscaterred phases to the wireless reader. 
The elements in our design which require power thus are the 2 RF switches and the 2 clock sources.
For switching at kHz frequencies, we observed that the power consumed by the 2 HMC544AE switches was almost similar to the static power consumption of 3.3 $\mu$W (the static current for switch at 3.3V voltage level is 0.5 $\mu$A~\cite{HMC544AE}). Although in our design we use a microcontroller to provide the clocks, by using low-powered oscillators, we can meet the clocking requirements with about 2 $\mu$W power budget ~\cite{shrivastava2012150nw}. Overall, the requirements are lesser than $10\mu$W which are modest enough to be supported by a RF energy harvesting circuit. In recent works, papers have even shown more than 50 $\mu$W power being harvested via RF signals, across 1cm of tissue~\cite{laiwalla2019distributed}.

\subsection{Wireless Reader Implementation}
\label{sec:usrp_impl}
\noindent The main task of the wireless reader is to transmit the OFDM waveform and periodically estimate the channel, so that phase changes at the shifted frequencies from the sensor can be read wirelessly. To perform the channel estimation, we utilize a 64 subcarrier, 12.5~MHz OFDM waveform.
We test this for both center frequency of 900~MHz and 2.4~GHz. We use separate antennas for transmission and reception, and use the same USRP N210 SDR~\cite{usrpref}\footnote{We can potentially use a COTS device as well as the wireless reader. Refer to Section~\ref{section:sdr_cots} for a brief discussion on the same} for both functions. Since the transmit and receive chains are on the same device, they are synchronized and will not show frequency/phase offsets. We emphasize here that the arduino clock is not synchronized with the other elements of the system since the force sensor is deployed as a separate entity.

We use a 320 sample long OFDM preamble padded with 400 zeros for the channel estimate. At the sampling rate used, this translates to fresh channel estimates every $T = \frac{720}{12.5\times10^6} = 60$ $\mu$s. Recall that to sense harmonics, the maximum harmonic frequency which can be sensed would be $|f_{\text{max}}| = \frac{1}{2T} \approx 8.7$~kHz due to the Nyquist Limit. We therefore chose our sensor clock frequencies to be 1,2~kHz, which would give modulation at 1,4~kHz, and falls comfortably within measurable limits. 


\begin{figure*}[t!]
  \centering
  \begin{subfigure}[t]{0.32\linewidth}%
  \centering
  \includegraphics[width=\linewidth]{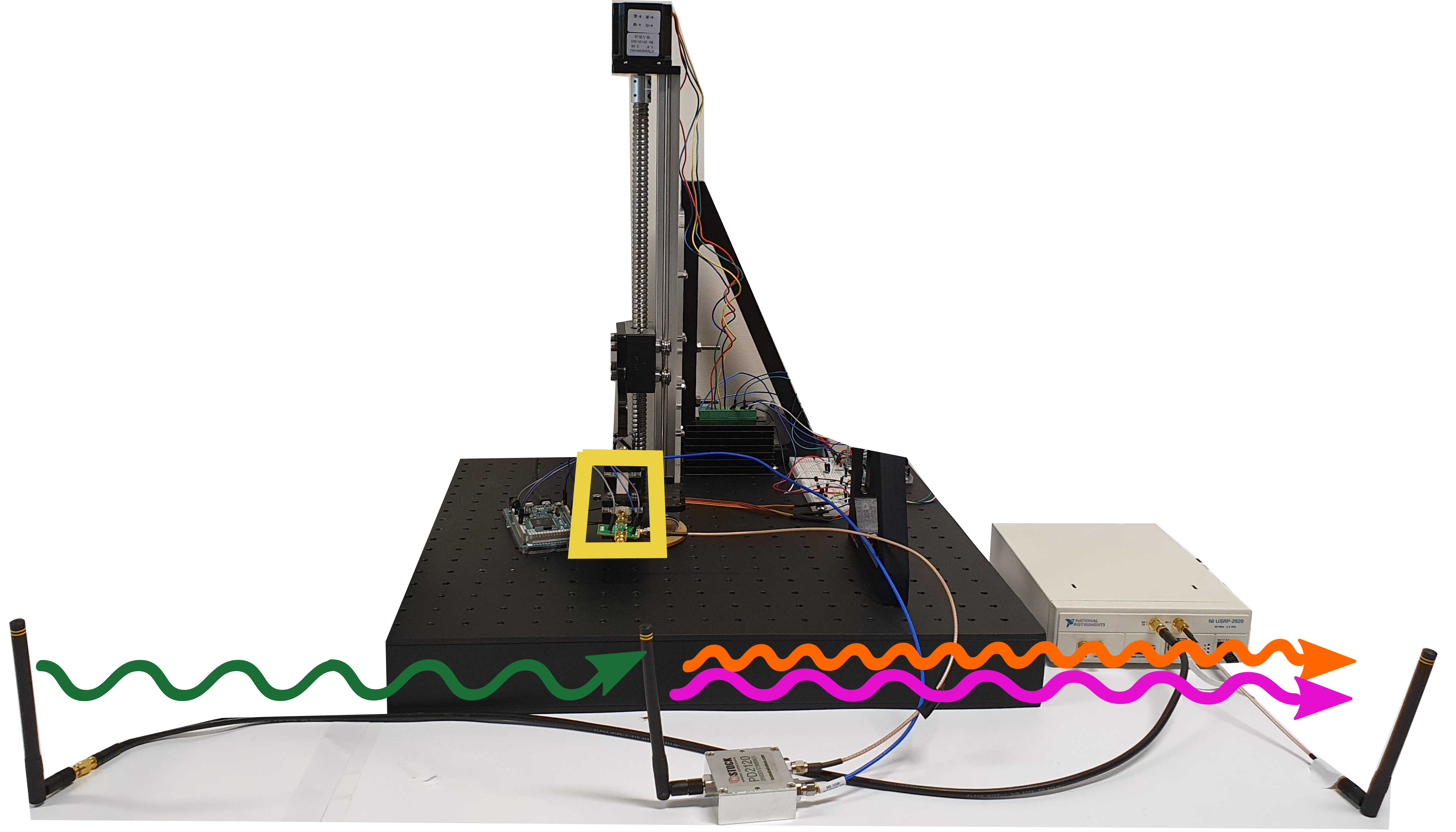}
  \put(-155,33){\setlength{\fboxsep}{1pt}\fcolorbox{black}{white}{\scriptsize{\shortstack[l]{TX}}}}
  \put(-148,32){\vector(0,-1){10}}
  \put(-12,32){\setlength{\fboxsep}{1pt}\fcolorbox{black}{white}{\scriptsize{\shortstack[l]{RX}}}}
  \put(-5,30){\vector(0,-1){10}}
  \put(-150,67){\setlength{\fboxsep}{1pt}\fcolorbox{black}{white}{\scriptsize{\shortstack[l]{Linear actuator\\ + indenter}}}}
  \put(-105,67){\vector(1,-1){15}}
  \put(-157,50){\setlength{\fboxsep}{1pt}\fcolorbox{black}{white}{\scriptsize{\shortstack[l]{Sensor on the\\ load cell platform}}}}
  \put(-105,53){{\vector(1,-1){15}}}
  \put(-140,2){\setlength{\fboxsep}{1pt}\fcolorbox{black}{white}{\scriptsize{\shortstack[l]{Sensor antenna}}}}
  \put(-95,5){{\vector(1,0){10}}}
  \caption{\textbf{Over the air wireless evaluation setup}}
  \label{fig:wireless_setup_general}
  \end{subfigure}
  \hfill
  \begin{subfigure}[t]{0.32\linewidth}%
  \centering
  \includegraphics[width=\linewidth]{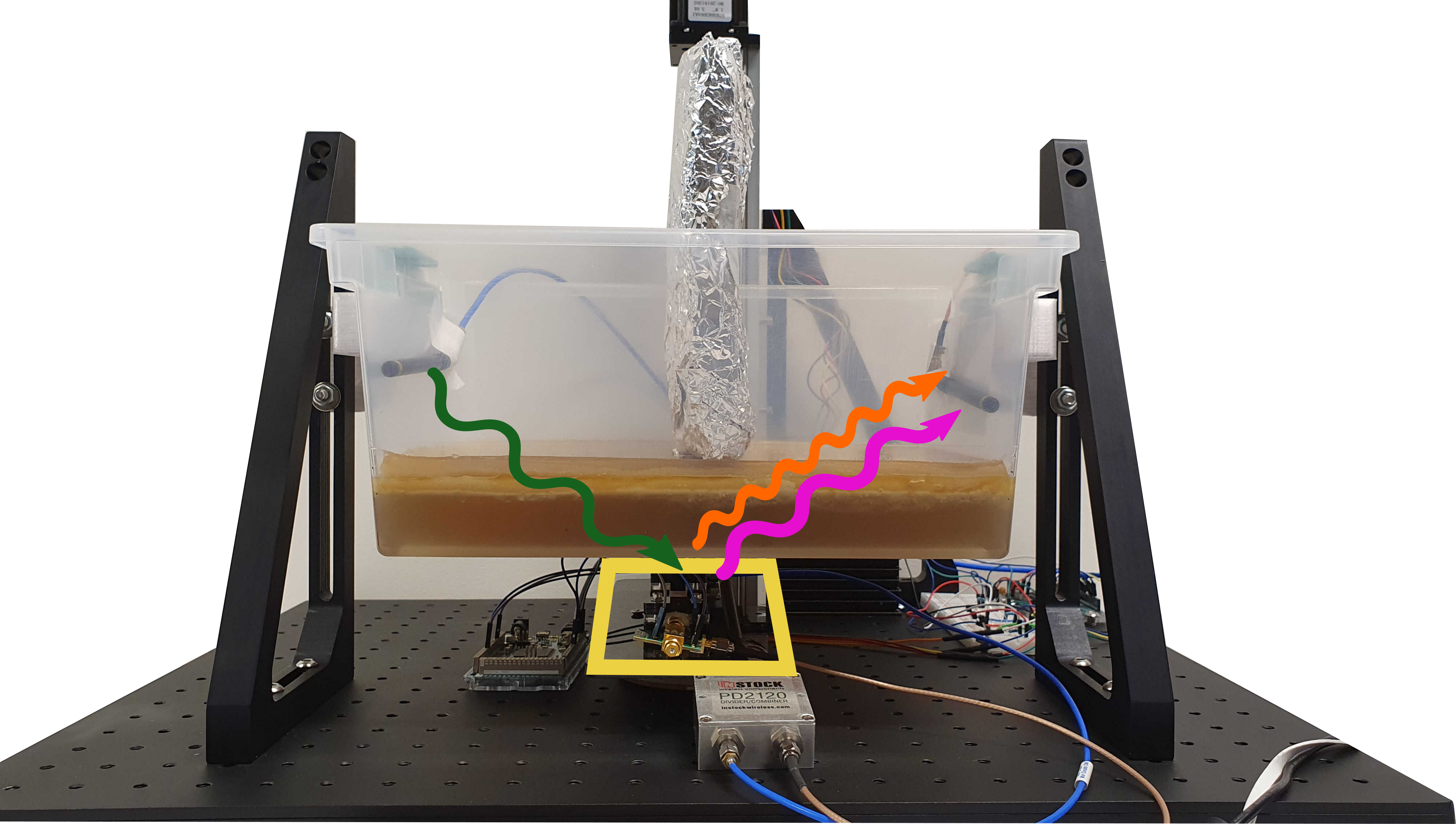}
  \put(-145,5){\setlength{\fboxsep}{1pt}\fcolorbox{black}{white}{\scriptsize{Sensor with antenna}}}
  \put(-100,10){\textcolor{white}{\vector(2,1){20}}}
  \put(-50,5){\setlength{\fboxsep}{1pt}\fcolorbox{black}{white}{\scriptsize{\shortstack[l]{Tissue\\phantom}}}}
  \put(-47.5,15){\textcolor{white}{\vector(-1,2){10}}}
  \put(-136,70){\setlength{\fboxsep}{1pt}\fcolorbox{black}{white}{\scriptsize{\shortstack[l]{RF isolation}}}}
  \put(-100,70){\vector(1,0){20}}
  \caption{\textbf{Tissue phantom setup}}
  \label{fig:gelatin_experiment}
  \end{subfigure}
  \hfill
  \begin{subfigure}[t]{0.3\linewidth}%
  \centering
  \includegraphics[width=\linewidth]{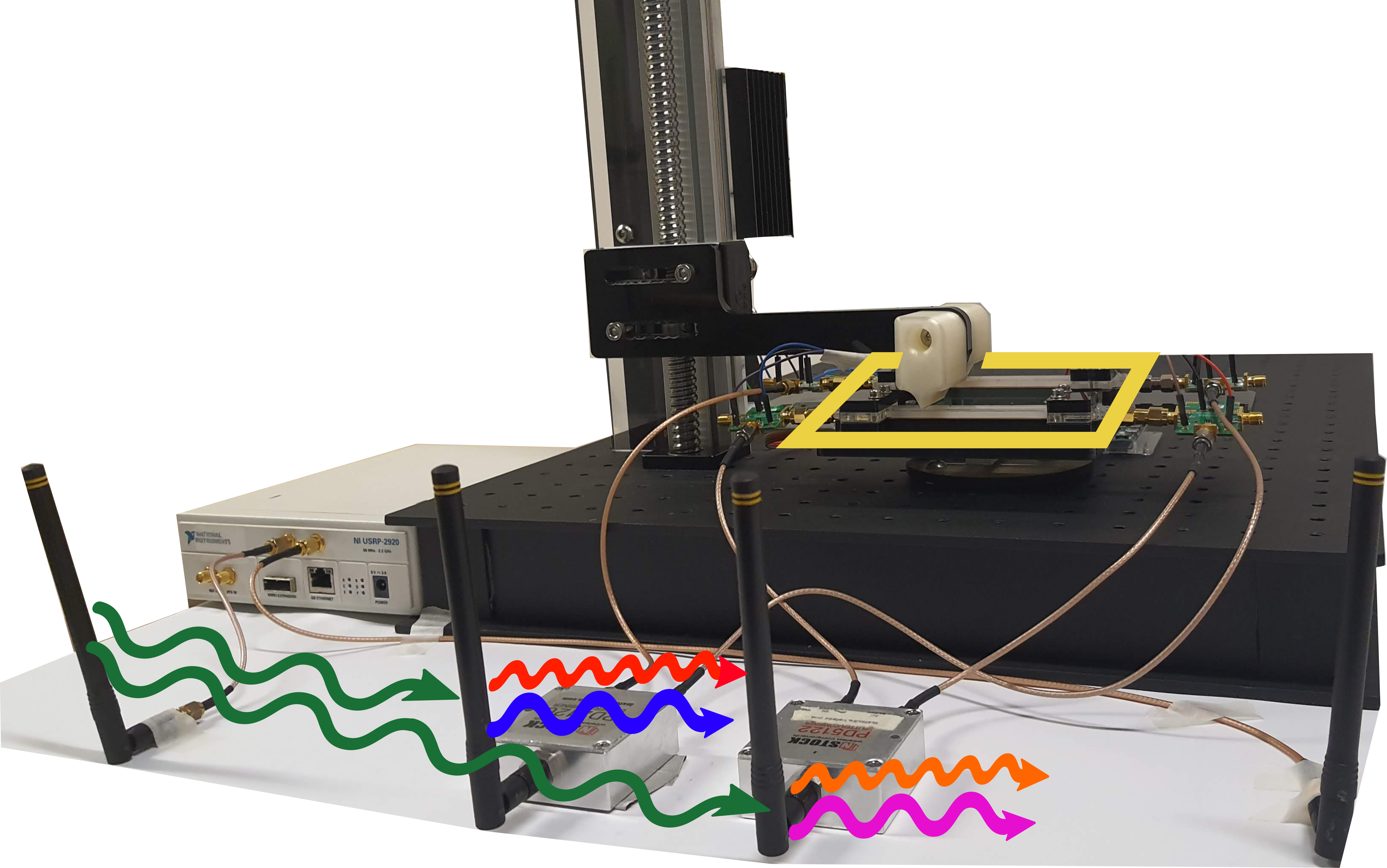}
  \put(-60,70){\setlength{\fboxsep}{1pt}\fcolorbox{black}{white}{\scriptsize{\shortstack[l]{Two sensors on the \\load cell platform}}}}
  \put(-35,67){\vector(0,-1){15}}
  \caption{\textbf{Multi Sensor experiment setup}}
  \label{fig:multisens_experiment}
  \end{subfigure}
  \caption{Different evaluation settings for \name's sensor prototype. In (a), the distance between TX-sensor antenna is 50cm, sensor antenna-RX is 50cm. In (b), the antenna and sensor placement ensures that propagation happens through the tissue phantom. In both (a), (b), green: excitation signal, orange,pink: freq. shifted backscattered signal. We also show capability to read from multiple sensors via the setup in (c), here, the red/blue waves represent a different set of frequency shifts for the second sensor}
  \label{fig:eval_settings}
\end{figure*}
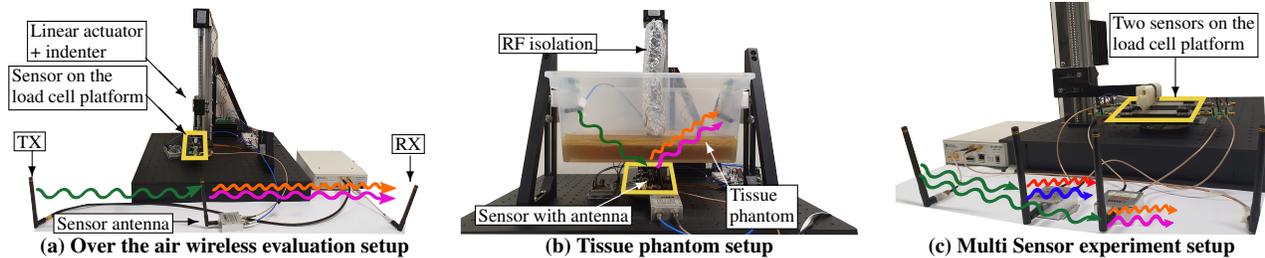

\begin{figure*}[t!]
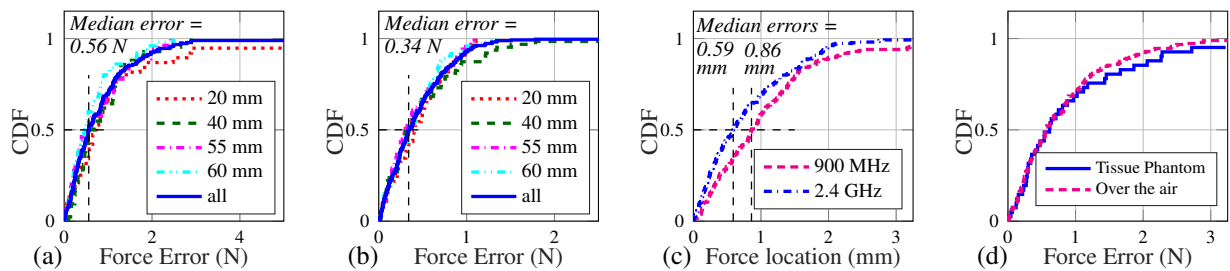

\centering
\begin{subfigure}[t]{0.24\linewidth}%
 {\input{results/cdf_900_mosaic}\label{fig:cdf_900}}
\put(-81,85){\footnotesize{\textit{\shortstack[l]{Median error =\\0.56~N}}}}
\put(-95,5){(a)}
\end{subfigure}%
\hfill
\begin{subfigure}[t]{0.24\linewidth}%
 {\input{results/cdf_2400_mosaic}\label{fig:cdf_2400}}
\put(-81,85){\footnotesize{\textit{\shortstack[l]{Median error =\\0.34~N}}}}
\put(-95,5){(b)}
\end{subfigure}%
\hfill
\begin{subfigure}[t]{0.24\linewidth}%
\input{results/force_loc_cdf_mosaic}
\put(-82,93){\footnotesize{\textit{Median errors =}}}
\put(-82,77){\footnotesize{\textit{\shortstack[l]{0.59\\mm}}}}
\put(-64,77){\footnotesize{\textit{\shortstack[l]{0.86\\mm}}}}
\put(-95,5){(c)}
\label{fig:force_loc_cdf}
\end{subfigure}%
\hfill
\begin{subfigure}[t]{0.24\linewidth}%
{\input{results/gelatin_cdf_mosaic}\label{fig:gelatin_cdf}}
\put(-95,5){(d)}
\label{fig:gelatin_tests}
\end{subfigure}%
\vspace{-1mm}
\caption{\textbf{Force CDF Plots}, 900 MHz (a), 2.4 GHz (b). For testing at individual points, the error is comparable to the combined CDF for which readings across the length are collected. Hence, the performance of the sensing strategy remains good throughout the sensor length. (c) shows the localization CDF for both the frequencies and (d) compares over the air performance with the tissue phantom setup. Similar CDF plots for both the setups show the sensing robustness when tested with the tissue phantoms}
\label{fig:force_cdf}
\vspace{1mm}
\end{figure*}

\section{Experimental Evaluation}
\vspace{0.5mm}
\noindent Armed with a sensor model to get from phases to force magnitude/location, as well as wireless reader and sensor implementation to enable backscatter sensing capabilities, we now evaluate the wireless performance of our sensor in different indoor environments.
The developed sensor model is first used to estimate the force magnitude and location exerted on the sensor, and this predicted force magnitude/location is compared to ground truth readings from the load cell and actuator position, as shown in Fig.~\ref{fig:wireless_setup_general}.
In addition, we plot empirical CDFs, which allow us to understand the accuracy of our sensing solution. 
Not restricting to over the air evaluations, we even evaluate our sensor when the wireless propagation occurs through tissue phantoms made to emulate human tissues. 
We also show the capability of reading forces from two sensors simultaneously.
Finally, we show that the force sensing works not only with the precision touches of the actuator, but can also detect the force and contact location when a human interacts with the sensor via finger touches.

\subsection{Wireless Performance Evaluation}

\noindent The first step in the wireless performance evaluation is to verify if the estimated force magnitudes and locations agree with the ground truth force-phase curves obtained via the load cell and the VNA setup. For this purpose, the setup illustrated in Fig.~\ref{fig:wireless_setup_general} is used\footnote{We also evaluate the performance of the sensing algorithm over a range of distances till 2m. The results are presented in Section~\ref{sec:dist_exp}}, with the sensor on top of a platform having load cell to give ground truth readings for the experiment, similar to the VNA experiment in Section~\ref{subsec:vna_meas}. Forces are applied between 0 and 8~N at 20, 40, 55 and 60~mm positions on the sensor. From Table~\ref{table:open_closed_loop_figures}, we can clearly see that wireless sensing is able to follow the VNA force-phase curves. Hence, this allows to validate the wireless implementation.


Using the estimated values of force magnitudes and force location to the ground truth, i.e. load cell readings and indenter location, we plot emperical CDFs to evaluate the wireless performance metrics. In Fig.~\ref{fig:force_cdf}a, Fig.~\ref{fig:force_cdf}b, we see that median error of force magnitude estimation of \paperTitle is 0.56~N when being read at 900 MHz, and 0.34~N when being read at 2.4~GHz. These results are satisfactory, since the errors are a fraction of the operating range of the sensor, which is approximately 8~N. One can observe that the error is lower at high frequency. Since higher frequency signals accumulate more phases for the same travelled distance, the required granularity for phase sensing is more relaxed, leading to lower error than sensing at low frequencies. Another observation from Fig.~\ref{fig:force_cdf}a, Fig.~\ref{fig:force_cdf}b, is that the sensor works uniformly across its length, i.e. error CDFs are similar when plotted for touching at individual locations with increasing magnitude of forces. 

Proceeding similarly, the median errors on the estimated force location is 0.86~mm at 900~MHz, and 0.59~mm at 2.4~GHz, as visible in Fig.~\ref{fig:force_cdf}c. Similar to force magnitude CDFs, performance is better at a higher frequency, since more phase change is accumulated per unit length at higher frequencies, enabling finer location estimation. These location results are satisfactory, with about 5 times higher accuracy than reported in recent work~\cite{pradhan2017rio,gao2018livetag}, where errors are in the order of magnitude of centimeters. The reasons for this improved performance are two-fold. To localize the contact location, \name correlates the extra separation between the shorting points caused by sensor bending in the action of a certain contact force. This correlation is enabled by the novel two-ended sensing strategy of \name.
This is fundamentally very different from the past contact location sensing approaches~\cite{pradhan2017rio,gao2018livetag}. Furthermore, this is supported by a wideband phase sensing algorithm (Section~\ref{Subsec:wideband}), which is capable of sensing these phases very accurately and robustly, unlike the previous works which used a narrowband RFID reader for the evaluations~\cite{pradhan2017rio}.  

\subsection{Testing with Tissue Phantoms}



\noindent We now assess the performance of our backscatter sensing strategy through human tissue. Propagation through human tissues necessitates using 900 MHz over 2.4GHz, as frequencies higher than 1 GHz are severely attenuated in such environments~\cite{gupta2003towards,dove2014analysis}. 
Wireless signals undergo huge losses when they propagate through human tissue, since these tissues are typically materials with high dielectric constants (with $\epsilon_r>10$)~\cite{vasisht2018body}. 
Further, the propagation is hampered via refraction and total internal propagation effects, which exacerbate the losses. 
Thus, to sense the robustness of our strategy with these impairments, we use the setup visible in Fig.~\ref{fig:force_cdf}d. 
It consists of a tissue phantom composed of three layers (muscle, fat and skin, and thickness of 25, 10 and 2~mm, respectively), with dielectric properties selected to mimic human tissue properties, as in~\cite{5523906}.


During these experiments, we observe that there was around 110~dB two-way backscatter loss from the TX-sensor and sensor-RX, for center frequency 900 MHz, when communicating through the tissue phantom. However, the direct path TX to RX signal had about 10-15~dB loss. The dynamic range of the USRP SDR we use was around 60~dB, because of which we can't decode the weak backscattered signal under the presence of the much stronger direct path signal. Hence, for these experiments, we isolated the TX and RX with a metal plate for this experiment. Because of the metal plate, the direct path loss increased to about 60~dB, which allowed us to decode the 50~dB lower backscattered signal at the receiver using the 60~dB dynamic range ADC of the USRP. For this experiment, we apply contact force at 60~mm on the sensor. We obtain similar performance as with the over-the-air tests, with the median force error increasing slightly from 0.56~N to 0.62~N (Fig.~\ref{fig:force_cdf}d). These results demonstrate the robustness of \paperTitle's wireless capabilities, since the sensing algorithm was able to decode force readings from a weaker signal trough the tissue phantom. In future works, the metal blockage can be replaced by self-interference canceling strategies, however, this is beyond the scope of this paper.



\subsection{Multi-sensor experiments}
\noindent We also evaluate the capability of \name to sense from multiple sensors simultaneously. The setup here consists of two sensors placed on a platform, and we use a custom designed indenture with the actuator in order to press on the two sensors simultaneously (Fig.~\ref{fig:multisens_experiment}). A load cell is attached below the platform to measure the combined forces acting on the platform (Fig.~\ref{fig:multi_sens_res}), whereas via wireless sensing from the two sensors we can estimate $F_1, F_2$ individually. In order to have separate identities for the two ends of the other sensor, we modulate via 1400, 2800 Hz duty cycled waves (visually illustrated via red, blue waves Fig.~\ref{fig:multisens_experiment}, Fig~\ref{fig:multi_sens_res}). 

By reading at these frequencies, we can wirelessly obtain estimates $\hat{F_1}, \hat{F_2}$ of $F_1,F_2$. Because the load cell measures $F_1+F_2$, we expect that adding these two estimates should allow us to compare against the ground truth load cell readings. The added estimates are expected to give a median error of $1.12N$, since one estimate from the sensor comes with a median error of $0.56N$ at 900 MHz, as profiled by the CDF plots (Fig.~\ref{fig:force_cdf}). Thus, we plot $F_1+F_2 \pm 1.12N$ as the blue shaded region as the expected median performance of the sensor to sense the added force.  Indeed, we see the added up estimates respecting the median error by being confined inside the median error region (Fig.~\ref{fig:multi_sens_res}). 

\subsection{Getting More Than Finger Touch: Measuring Fingertip Forces}

\noindent We now motivate a UI use case, which has the potential to improve and change the way users interact with digital devices. For this purpose, we select a center frequency of 2.4 GHz for our sensor, which is well-adapted to Wi-Fi and Bluetooth devices.
To assess the relevance of our sensor for such applications, we use the fingertip, instead of the actuated indenture, to press the sensor with varying force levels.
An operator presses the sensor at the 60~mm location. We plot the force readings from the load cell in real time, and use this real-time plot to give the user visual cues to settle in to some force level. Then, we estimate these force levels using \paperTitle to evaluate if it can support these sensing capabilities. 
\begin{figure}[t!]
\centering
\begin{subfigure}[t]{0.47\linewidth}
\centering
\includegraphics[width=\linewidth]{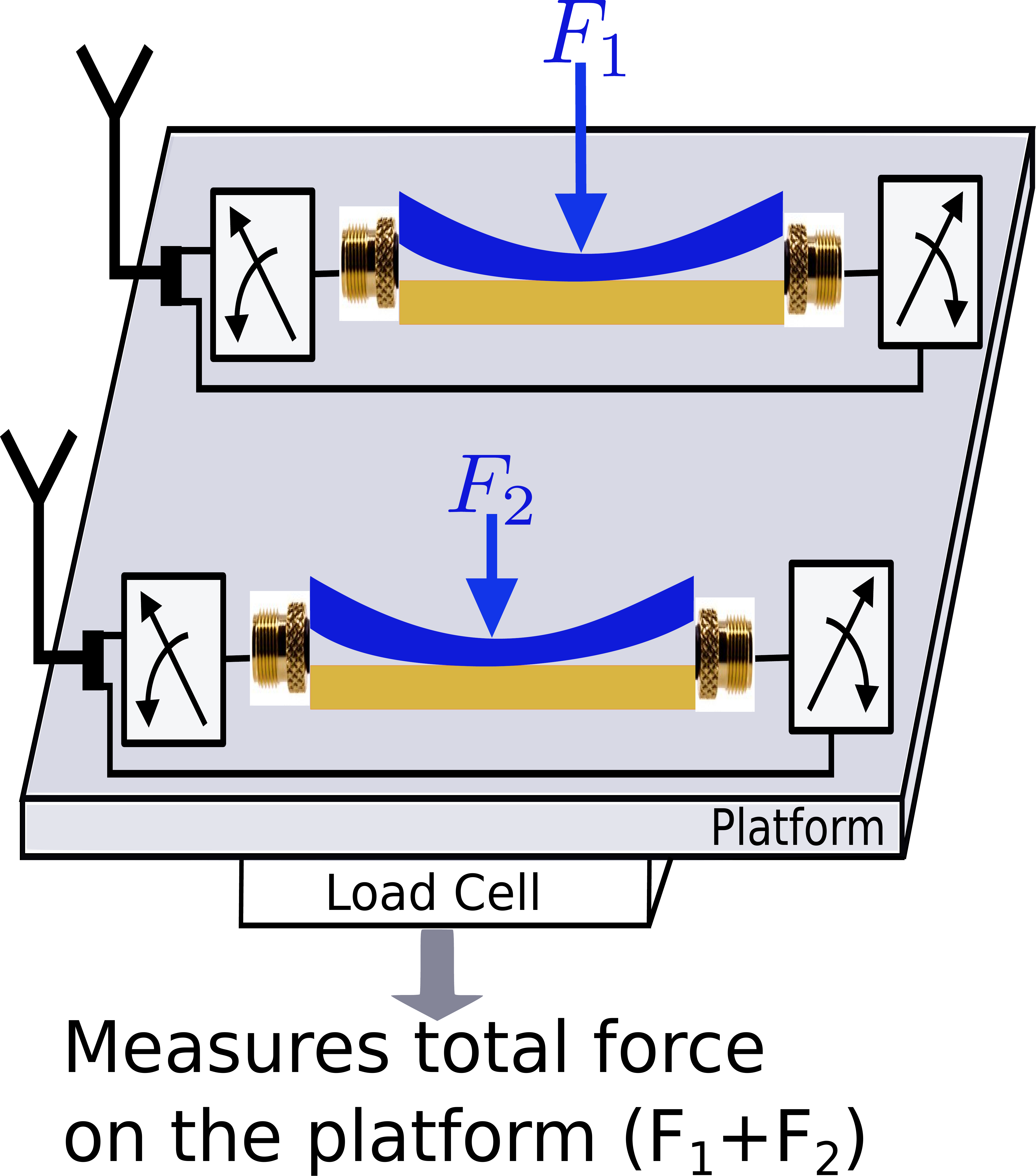}
\end{subfigure}
\begin{subfigure}[t]{0.47\linewidth}
\centering
{
%
%
\definecolor{mycolor1}{rgb}{0.00000,0.44700,0.74100}%
\definecolor{mycolor2}{rgb}{0.85000,0.32500,0.09800}%
\definecolor{mycolor3}{rgb}{0.92900,0.69400,0.12500}%
\definecolor{mycolor4}{rgb}{0.49400,0.18400,0.55600}%
\begin{tikzpicture}

\begin{axis}[%
width=1.1in,
height=1.25in,
at={(1.297in,1.093in)},
scale only axis,
xmin=0,
xmax=20,
xlabel={Time (samples)},
ymin=2,
ymax=22,
ylabel style={at={(axis description cs:-0.1,0.425)},anchor=south},
ylabel={Force (N)},
axis background/.style={fill=white},
xmajorgrids,
ymajorgrids,
legend columns=2,
legend style={at={(0.015,0.875)},anchor=west, nodes={scale=0.5, transform shape}, legend cell align=left, align=left, draw=white!15!black},
]
\addplot [color=mycolor1, line width=1.5pt]
  table[row sep=crcr]{%
1	7.473528\\
2	8.147572\\
3	8.647764\\
4	9.124044\\
5	9.582292\\
6	10.068568\\
7	10.518388\\
8	11.431944\\
9	11.832274\\
10	12.305908\\
11	12.699378\\
12	13.136654\\
13	13.526204\\
14	13.981414\\
15	14.367534\\
16	14.786582\\
17	15.253062\\
18	15.50404\\
19	16.087924\\
};
\addlegendentry{$F_1+F_2$}

\addplot [color=mycolor2, line width=1.5pt, mark size=1.5pt, mark=asterisk, mark options={solid, mycolor2}]
  table[row sep=crcr]{%
1	6.9\\
2	8.21866529281538\\
3	9.10331041452525\\
4	10.3081638614544\\
5	11.2788291558755\\
6	10.9958966445395\\
7	9.76978925564578\\
8	11.4681536640849\\
9	11.1812590599\\
10	12.5159448665251\\
11	12.3120594448338\\
12	13.0976221107944\\
13	13.1474651929752\\
14	12.7579659883767\\
15	13.7157686636083\\
16	13.3725830918162\\
17	15.1175636657004\\
18	16.0396753255309\\
};
\addlegendentry{$\hat{F_1}+\hat{F_2}$}

\addplot [color=mycolor3, line width=1.5pt, mark size=1.5pt, mark=diamond, mark options={solid, mycolor3}]
  table[row sep=crcr]{%
1	4.6\\
2	5.06839902338998\\
3	5.7886639957815\\
4	6.58324761457635\\
5	6.93672940383674\\
6	6.01210877583565\\
7	5.58021378231875\\
8	6.17470892944864\\
9	5.96606643151751\\
10	6.96769528613418\\
11	6.62884454780506\\
12	6.54891599818323\\
13	6.69070300383753\\
14	6.68319815091491\\
15	7.49684442959675\\
16	6.85739567292224\\
17	7.46662072608399\\
18	8.0071173633948\\
};
\addlegendentry{$\hat{F_1}$}

\addplot [color=mycolor4, line width=1.5pt, mark size=1.5pt, mark=triangle, mark options={solid, mycolor4}]
  table[row sep=crcr]{%
1	2.3\\
2	3.1502662694254\\
3	3.31464641874375\\
4	3.72491624687802\\
5	4.34209975203875\\
6	4.98378786870386\\
7	4.18957547332702\\
8	5.29344473463628\\
9	5.21519262838246\\
10	5.54824958039096\\
11	5.68321489702875\\
12	6.54870611261121\\
13	6.45676218913767\\
14	6.07476783746175\\
15	6.21892423401154\\
16	6.51518741889397\\
17	7.65094293961646\\
18	8.03255796213614\\
};
\addlegendentry{$\hat{F_2}$}

\addplot[area legend, draw=none, fill=blue, fill opacity=0.3]
table[row sep=crcr] {%
x	y\\
1	6.353528\\
1	8.593528\\
19	17.207924\\
19	14.967924\\
}--cycle;
\end{axis}

\end{tikzpicture}%
}
\end{subfigure}
\caption{Sensing forces simultaneously from two sensors}
\label{fig:multi_sens_res}
\end{figure}
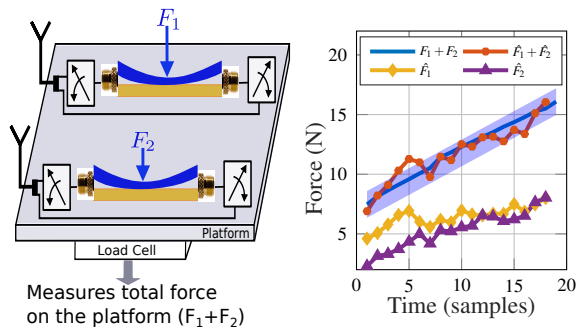

\begin{figure}[t!]
\begin{subfigure}[t]{0.15\textwidth}
\centering
 {
%
%
\definecolor{mycolor1}{rgb}{0.00000,0.44700,0.74100}%
\begin{tikzpicture}

\begin{axis}[%
width=0.75in,
height=1.2in,
at={(1.297in,1.093in)},
scale only axis,
xmin=0,
xmax=80,
xlabel={Location (mm)},
ymin=0,
ymax=0.8,
ylabel style={at={(axis description cs:-0.18,0.425)},anchor=south},
ylabel={Probability},
axis background/.style={fill=white},
xmajorgrids,
ymajorgrids,
legend style={at={(-0.1,0.15)},anchor=west, nodes={scale=0.5, transform shape}, legend cell align=left, align=left, draw=white!15!black}
]
\addplot[ybar interval, fill=mycolor1, fill opacity=0.6, draw=black, area legend] table[row sep=crcr] {%
x	y\\
0	0\\
10	0.00666666666666667\\
20	0.00666666666666667\\
30	0.00666666666666667\\
40	0.0666666666666667\\
50	0.7\\
60	0.206666666666667\\
70	0.00666666666666667\\
80	0\\
90	0\\
100	0\\
};

\end{axis}
\end{tikzpicture}%
}\hfill
 \caption{Finger touch location histogram.}
 \label{fig:finger_touch_loc}
\end{subfigure}
\hfill
\begin{subfigure}[t]{0.35\textwidth}
\centering
 {\input{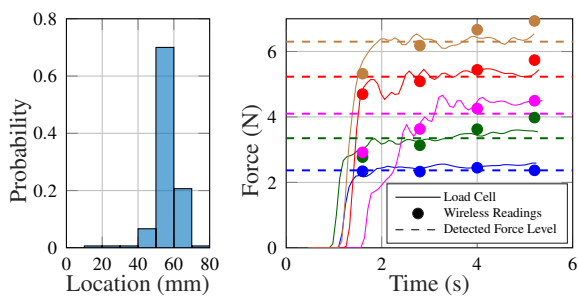}}
 \caption{Finger force level readings vs time.}
 \label{fig:finger_touch_mag}
\end{subfigure}
\caption{Wireless Sensing results for pressing at 60~mm with increasing force levels via a fingertip. From (a) we see that all the touch interactions at 60~mm $\pm 20$mm were classified correctly, as the sensor was pressed with a finite-width fingertip (about 15-20 mm~\cite{dandekar20033,young2020compensating}). From (b), it can be seen that \paperTitle was able to estimate increasing force levels accurately}
\label{fig:finger_touch}
\end{figure}

Fig.~\ref{fig:finger_touch} shows the evaluation results of \paperTitle's experiments. From Fig.~\ref{fig:finger_touch_loc} one can see that the sensor could accurately detect the pressing location, which was 60~mm, with sufficient accuracy, considering the fact that a typical human fingertip has a width and thickness of approximately 15-20~mm~\cite{young2020compensating,dandekar20033}. That is, even though \name's location sensing had sub-mm location sensing accuracy, now the error source will most likely be coming from the uncertainity of how operators press the sensor with their finite width fingertips, instead of the precise point-pressing feature of the actuator before. Since most of the readings are clustered $\pm 20$mm near the visual cue of $60$mm given to the operators, \name does operate under the practical limit of the experimental setting due to finite width of the operators.

Further, in Fig.~\ref{fig:finger_touch_mag}, we see how \paperTitle is able to get more than just binary touch sensing results. Not only can \paperTitle detect the point where a finger touched the sensor, going one step ahead, \paperTitle is able to detect the force profiles of the touch interactions as well, which motivates much improved UI use cases by getting more than just touch/no touch information.

\fi


\section{Discussions and Applications}
\noindent The most natural usecase for such wireless haptic feedback lies in surgical robots and tools. 
Human hands are extremely dexterous, and provide unparalleled sensory feedback which enable very precise operations required for surgery. However, we need tools and robots to emulate the human hands when direct operation is not possible, such as during minimally invasive surgical operations. Ideally surgeons should receive haptic feedback from the tools/robots they are operating, which would require information of both force magnitude and contact location. 
However, such haptic feedback is generally not available in practice, and one reason has been that force sensing modalities are still not evolved enough to support these applications~\cite{haouchine2018vision,aviles2015sensorless,dargahi2012tactile,okamura2009haptic,reiley2008effects}.
Loss of haptic feedback increases the training time for surgeons, increases risk of surgical errors, and hinders the closed-loop operation for robot assisted surgeries~\cite{kothari2002training,zhou2012effect,yip2016model,tholey2005force}.

The current form factor and sensor interface hinders direct use of \paperTitle's sensor in more complex surgical tasks requiring force feedback, such as cardiac ablation~\cite{yip2016model} or pre-retinal membrane peeling~\cite{francone2019effect}. However, the sensor can help solve a major problem in laparoscopy known as the fulcrum effect~\cite{nisky2012perception}. The fulcrum effect is caused due to lever effect caused by contact forces between the body and surgical tool, at the entry point of the surgical incision. Due to lack of feedback on both the magnitude of force and location, the tool tend to slip, which causes risks of tissue damage. A laparoscopic surgical tool augmented with a \paperTitle sensor to determine and localize the contact force can prevent this fulcrum effect since the surgeon can do a closed loop correction based on this haptic feedback.

Apart from surgical applications, sensing contact force and location can be extremely useful for robotic tasks which require a manipulator/gripper. Robotic manipulators need this haptic feedback to determine how firmly they have grasped a particular object~\cite{billard2019trends, deng2020grasping}. People have attempted doing this via vision induced haptics~\cite{yuan2017gelsight,she2020exoskeleton}, however, these methods typically require computationally intensive algorithms and fail to meet the required temporal rate of feedback required to determine if the grasp of the object is loosening and slipping~\cite{chen2018tactile}. However, since wireless sensing is not bound to such issues, and can be made near real-time. Thus, such sensors can be used for direct and low-latency haptic feedback to improve robotic manipulation operations.

\noindent Alongside the robotics centered applications, force sensing can have many latent applications in the next generation interfaces for HCI/AR-VR. Smart surfaces have been an active area of research, with touch sensing touted to a game changer for ubiquitous computing~\cite{ishii2008tangible,zhang2019sozu,gao2019livetagGetmobile}. Force sensing will add more depth to these touch sensing solutions, and can lead to some unforeseen applications.


\section{Future Work}
%

\noindent \textbf{Extending to 2-D continuum}: The current sensor prototype of \paperTitle consists of sensing on a 1-D continuum. To extend this sensing to a 2-D continuum, we can deploy multiple \paperTitle sensors placed next to each other. Hence, by reading phase changes from multiple \paperTitle sensors, we can infer the location and contact force magnitude on the 2-D continuum spanned by these multiple sensors. A hindering factor to this 2-D extension is how to address multiple touch points simultaneously, which will be explored in future works.

\noindent \textbf{Reducing the form factor:} \paperTitle is the first work which presents such a low-powered sensor, and thus naturally leads the way to realize a battery-free haptic feedback. The current sensor prototype of \paperTitle is 80~mm long, and about 10~mm thick. With the current form factor, the sensor is not directly applicable for some of the medical applications which need smaller sensors. The sensors can get to the correct form-factor requirements by designing integrated circuits, antenna and the sensor fabrication. 
To make the sensor prototype more flexible, we will explore new fabrication strategies like flexible PCB printing and creating custom RF connectors.




\section{Related Work}

\noindent Force sensors have been developed using a variety of transduction mechanisms, such as capacitive, piezoresistive, piezoelectric, optical, magnetic, and inductive~\cite{chi2018recent}. There are a number of tradeoffs among the various mechanisms, including, for example, sensitivity, spatial resolution, accuracy, power consumption, and size. To meet the requirements of many emerging systems, particularly those where it may be difficult to have a physical wired connection to the sensors, many researchers have been investigating the creation of wireless sensors.

\textbf{Wireless force sensors:} 
A number of wireless capacitive force sensors that leverage a change in capacitance due to deformation have been recently developed. For example, a flexible capacitive sensor was created for wirelessly measuring strain in tires~\cite{matsuzaki2007wireless}, and a capacitive textile sensor was developed for wireless respiratory monitoring ~\cite{hoffmann2010respiratory}. While the capacitive sensing paradigm can work well for a number of force sensing applications, it is not naturally compatible with wireless sensing. In order to wirelessly transmit force information obtained through capacitive sensing, additional hardware and circuits are needed, complicating the design. 

Inductor-capacitor (LC) wireless sensors are passive devices that can remotely sense a number of parameters, including pressure. The working principle of these sensors is based on changes in the capacitance that causes a shift in the LC resonant frequency, which can be wirelessly measured~\cite{li2015review, huang2016lc}.
A number of these LC sensors have been developed for applications like 
monitoring of pressures during arterial blood flow~\cite{boutry2019biodegradable}, and the measure of finger tip forces during athletic activities~\cite{nie2018droplet}. However, the resonance frequency of these sensors is in the range of a few hundred kHz to a few MHz~\cite{huang2016lc}, which makes wireless sensing difficult. As a consequence, these sensors suffer from short interrogation distances in the range of a few centimeters~\cite{bau2018contactless,huang2016lc}.

There has also been a large body of research on strain sensors~\cite{li2018miniature,yi2011passive,thai2011design,humphries2012passive,teng2019soft}. In strain sensing, instead of sensing the normal transversal force, the longitudinal force is sensed. 
Longitudinal force has a tendency to stretch and elongate the object it is acting upon, hence these sensors estimate the change in the length to infer strain. 
Thus, most of the wireless strain sensors exploit the shifts in resonant frequencies to sense strain. 
That is, to infer strain, a wireless reader evaluates signal strength at multiple frequencies, to estimate the resonant frequency, where a notch will form in the signal strength measurements. 
It is well known that signal strength is a fickle quantity easily corrupted by multipath. 
Indeed, most of these works show evaluations in a controlled, anechoic environment, and the technology has not been found to be robust to static multipath~\cite{yi2011passive}.

\textbf{Backscatter sensing systems:} 
Recent advancements in `backscatter sensing' has enabled the creation of passive, battery-free touch interfaces. Touch sensing has been a well explored use case of RFID-based sensing~\cite{marquardt2010rethinking,sample2009capacitive,simon2014adding,schmidt2000enabling,gao2018livetag,pradhan2017rio,li2016paperid,li2015idsense}. IDSense~\cite{li2015idsense} utilized the fact that reflected RSS and phase change in a unique way when the RFID chip is touched, and following up on this PaperId~\cite{li2016paperid} even gave a simple manufacturing method by which one could simply use an inkjet printer to manufacture these RFID tags and augment everyday objects with touch interactions. RIO~\cite{pradhan2017rio} further explored the touch to reflected signal phase mapping to extend touch sensing to multiple RFID tags by utilizing mutual coupling effects, and extended the design further to use custom designed, application specific RFID tags. Livetag~\cite{gao2018livetag} presented a similar touch sensing system showing how to sense these touch interactions using Wi-Fi based readers, instead of relying on expensive, dedicated RFID readers used by earlier works. However, none of these systems could sense force magnitude and were limited to sensing just the position where the tag was being touched in order to sense simple gestures/sliding movements etc.

\section{Acknowledgements}
\noindent We thank anonymous reviewers and our shepherd, Prof. Haitham Hassanieh, for their insightful feedback. We also thank members of WCSNG, UCSD for their feedback throughout the process, as well as analog devices for providing us HMC544AE samples. This research was supported in part by the National Science Founda-tion under Grant 1935329.
\bibliographystyle{unsrt}
\balance
\bibliography{main}
\label{lastpage}
\section{Appendix}
\label{section:appendix}
\subsection{Replacing SDR with COTS}
\label{section:sdr_cots}
We can read the sensor using a COTS-WiFi implementation which can provide CSI, like Quantenna~\cite{quantenna-ds}. With COTS devices, periodic channel estimates can be obtained with similar time latency as compared to our implementation using SDRs. In our implementation, $T=60\mu s$ (Section \ref{sec:usrp_impl}), which is reasonably higher than packet sizes of $12\mu s$ achievable by 1 Gbps WiFi systems \cite{magistretti2011wifi}.  
A potential issue could be MAC overheads, as also cited in \cite{magistretti2011wifi}, however, we can alter the packet structure slightly to avoid backoffs which can mitigate against the MAC overheads.
However, with COTS devices, we will have to deal with
 CFO effects in the measured channel response.
 
In our implementation on USRP, we had TX and RX sharing the same RF chain. With COTS devices like quantenna, we will have TX and RX as separate devices which might not be able to share a clock, thus leading to frequency and phase offsets.
Although \name design is robust to phase offsets due to our differential phase sensing approach, we would need to counter CFO.
To counter the CFO effects, we can use the fact that CFO will remain same for both the direct path between TX and RX, and the reflected signal from the sensor. To do so, we can cnsider a differential sensing approach by caluclating phase relative to the direct path, and similar approaches have been explored to do so in past work \cite{kotaru2017position,xie2019md}

\subsection{HFSS Simulations}
\label{section:hfss_sims}
For an air substrate microstrip transmission line, we have the following equation which governs the impedance matching, $Z = 60\text{ln}\left[\frac{6h}{w} + \sqrt{1+\left(\frac{2h}{w}\right)^2}\right]$, where $h$ is the vertical separation between signal and ground trace, and $w$ is the width of the signal trace \cite{steer2019microwave}. 
Setting $Z=50~\Omega$ in the above equation, gives us the operating $\frac{h}{w}$ ratio to be approximately $5:1$. 
In order to interface SMA connectors to the air-substrate microstrip line designed, we have to increase the width of the ground trace so that the bottom legs of the SMA connector can be soldered directly to the ground trace.

\begin{figure}[H]
\begin{subfigure}[t]{0.15\textwidth}
\centering
 {\includegraphics[width=\linewidth]{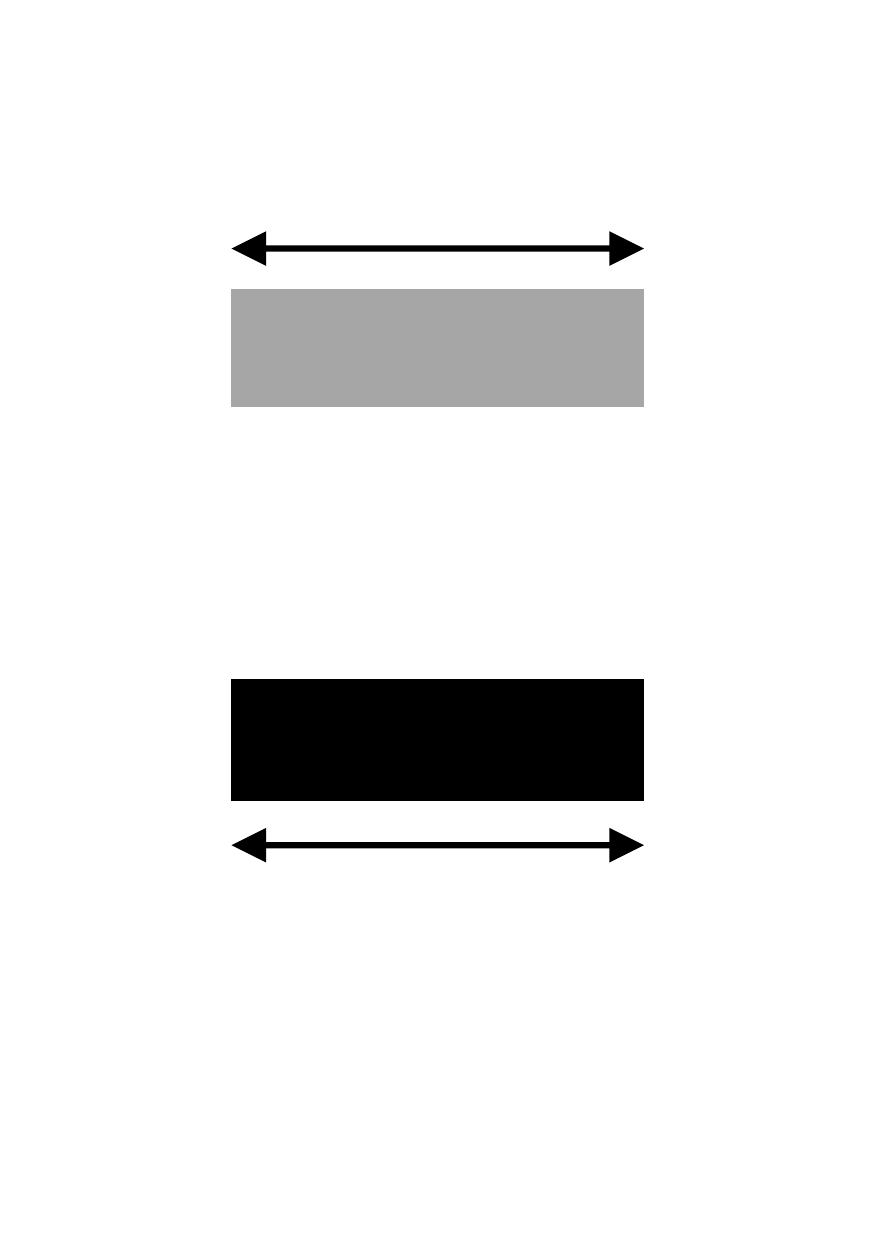}\label{fig:s11_mag_sim1}}
 \caption{Signal (gray) and Ground (black) traces\vspace{10pt}}
\end{subfigure}
\begin{subfigure}[t]{0.35\textwidth}
\centering
 {
%
%
\definecolor{mycolor1}{rgb}{0.00000,0.44700,0.74100}%
\definecolor{mycolor2}{rgb}{0.85000,0.32500,0.09800}%
\begin{tikzpicture}

\begin{axis}[%
width=1.4in,
height=1in,
at={(1.297in,1.093in)},
scale only axis,
xmin=1,
xmax=9,
xlabel={Height to Width Ratio},
ymin=-50,
ymax=19,
ylabel style={at={(axis description cs:-0.25,0.425)},anchor=south},
ylabel={S11 Magnitude (dB)},
axis background/.style={fill=white},
xmajorgrids,
ymajorgrids,
legend style={at={(0.4,0.83)},anchor=west, nodes={scale=0.75, transform shape}, legend cell align=left, align=left, draw=white!15!black}
]
\addplot [color=blue, line width=1.5pt]
  table[row sep=crcr]{%
8.33333333333334	-13.6147325113349\\
8.06451612903226	-14.7410072673383\\
7.8125	-15.6860425355949\\
7.57575757575758	-16.3687980509719\\
7.35294117647059	-17.7670630639589\\
7.14285714285715	-18.3916833895986\\
6.94444444444444	-19.9785722909735\\
6.75675675675676	-20.8001900792638\\
6.57894736842105	-22.7864476391003\\
6.41025641025641	-23.637906167676\\
6.25	-26.8210303742223\\
6.09756097560976	-28.7196095924292\\
5.95238095238096	-32.9219061925178\\
5.81395348837209	-34.0865844727597\\
5.68181818181818	-40.3591864816056\\
5.55555555555556	-37.9850648394873\\
5.43478260869566	-34.9282275378093\\
5.31914893617022	-29.7942042984264\\
5.20833333333334	-27.6021303172978\\
5.10204081632653	-26.1894864380822\\
5	-23.7441829389957\\
4.90196078431373	-22.4490717306518\\
4.80769230769231	-21.0975652120871\\
4.71698113207547	-20.2721979153779\\
4.62962962962963	-19.4415345032783\\
4.54545454545455	-19.006648061698\\
4.46428571428572	-18.2191116187994\\
4.3859649122807	-17.4208915154509\\
4.31034482758621	-17.2066129106662\\
4.23728813559322	-16.643348457733\\
4.16666666666666	-15.9714378399739\\
4.09836065573771	-15.4550094121872\\
4.03225806451613	-15.4397325372117\\
3.96825396825397	-14.7640553304681\\
3.90625	-14.2558162399323\\
3.84615384615385	-14.5568499918275\\
3.78787878787879	-13.946457587339\\
3.73134328358209	-13.709043294618\\
3.6764705882353	-13.3056936698603\\
3.6231884057971	-13.0517115978647\\
3.57142857142857	-13.0770019076594\\
3.52112676056338	-12.6511357428221\\
3.47222222222222	-12.5488328341039\\
3.42465753424658	-12.2335567958102\\
3.37837837837838	-12.0211183828896\\
3.33333333333334	-12.0433131909801\\
3.28947368421053	-11.4864854568299\\
3.24675324675324	-11.338497635608\\
3.2051282051282	-11.1356871112997\\
3.16455696202532	-11.3868229567886\\
3.125	-10.738874061207\\
3.08641975308642	-10.8121909993933\\
3.04878048780488	-10.386502402471\\
3.01204819277108	-10.4062247476204\\
2.97619047619047	-10.2456497150864\\
2.94117647058823	-10.2527681991737\\
2.90697674418605	-10.1219029198396\\
2.87356321839081	-9.96997743352419\\
2.84090909090909	-9.61643973672369\\
2.80898876404495	-9.78914620151287\\
2.77777777777778	-9.4270602184779\\
2.74725274725274	-9.28091565475905\\
2.71739130434783	-9.32169489178764\\
2.68817204301075	-9.1904927625861\\
2.6595744680851	-9.04873960005765\\
2.63157894736842	-8.82877488381512\\
2.60416666666666	-8.97502924475654\\
2.57731958762886	-8.6939347349103\\
2.55102040816327	-8.62380476402116\\
2.52525252525253	-8.61004410460789\\
2.5	-8.38291170575455\\
2.47524752475248	-8.28996995700327\\
2.45098039215686	-8.20650597302189\\
2.42718446601942	-8.32371352219798\\
2.40384615384615	-8.02044275273144\\
2.38095238095238	-7.9277140165576\\
2.35849056603774	-7.90220952769207\\
2.33644859813084	-7.86063980934673\\
2.31481481481482	-7.83477815413021\\
2.29357798165137	-7.89374731786168\\
2.27272727272727	-7.64995521053681\\
2.25225225225226	-7.61545136924383\\
2.23214285714285	-7.3824528573729\\
2.21238938053097	-7.56724318015497\\
2.19298245614035	-7.33864728969165\\
2.17391304347826	-7.31519160643714\\
2.1551724137931	-7.34435210153597\\
2.13675213675214	-7.11188837572397\\
2.11864406779661	-7.17857862600338\\
2.10084033613445	-6.97841130087403\\
2.08333333333334	-7.03840298006989\\
2.06611570247934	-6.97851796989627\\
2.04918032786885	-6.80479504931369\\
2.03252032520325	-6.91598162569378\\
2.01612903225806	-6.75373340243347\\
2	-6.80133179341767\\
1.98412698412699	-6.67606287044891\\
1.96850393700787	-6.59977333819414\\
1.953125	-6.63420632372869\\
1.93798449612403	-6.50107288452633\\
1.90839694656488	-6.3988638586067\\
1.89393939393939	-6.30471686720357\\
1.8796992481203	-6.32645176684709\\
1.86567164179105	-6.16577552923432\\
1.85185185185185	-6.15281189285891\\
1.83823529411764	-6.33781491361007\\
1.82481751824817	-6.10507826551382\\
1.81159420289855	-6.14928079055475\\
1.79856115107913	-5.95104840500326\\
1.78571428571428	-6.08329084407282\\
1.77304964539007	-6.10198433967801\\
1.76056338028169	-5.88842055076512\\
1.74825174825175	-6.23804857900498\\
1.73611111111111	-5.86904722396977\\
1.72413793103448	-5.76174134490113\\
1.71232876712329	-5.73419529786608\\
1.70068027210884	-5.68950787865055\\
1.68918918918919	-5.69139481530727\\
1.67785234899329	-5.7917645993641\\
1.66666666666666	-5.64495731273389\\
};
\addlegendentry{900 MHz}

\addplot [color=magenta, line width=1.5pt]
  table[row sep=crcr]{%
8.33333333333334	-19.1088596763856\\
8.06451612903226	-18.8158247491648\\
7.8125	-20.6623337773926\\
7.57575757575758	-21.3780840740369\\
7.35294117647059	-20.9456119114691\\
7.14285714285715	-21.3496580524922\\
6.94444444444444	-23.9609010588788\\
6.75675675675676	-24.5609175648807\\
6.57894736842105	-25.1059200902618\\
6.41025641025641	-24.090371079278\\
6.25	-27.5634507769579\\
6.09756097560976	-29.2806225436438\\
5.95238095238096	-29.3899509928589\\
5.81395348837209	-25.9949812731012\\
5.68181818181818	-29.9512070257845\\
5.55555555555556	-33.1841859415972\\
5.43478260869566	-36.2281063990529\\
5.31914893617022	-47.5179834564144\\
5.20833333333334	-46.8259920597216\\
5.10204081632653	-37.8221260045813\\
5	-43.1273690924893\\
4.90196078431373	-35.4218458525864\\
4.80769230769231	-32.4401811782262\\
4.71698113207547	-31.2929375002981\\
4.62962962962963	-29.5410975077574\\
4.54545454545455	-29.028732011624\\
4.46428571428572	-28.1648651515804\\
4.3859649122807	-28.6524079680994\\
4.31034482758621	-27.8364945755859\\
4.23728813559322	-26.2361930233866\\
4.16666666666666	-25.4371736035349\\
4.09836065573771	-24.4915985747989\\
3.96825396825397	-22.5299842980955\\
3.90625	-22.3503985526573\\
3.84615384615385	-22.0342526154324\\
3.78787878787879	-21.5744280730538\\
3.73134328358209	-21.301594742323\\
3.6764705882353	-21.2866351624252\\
3.6231884057971	-20.3892156956693\\
3.57142857142857	-20.5731347481125\\
3.52112676056338	-19.8847426625374\\
3.47222222222222	-20.5457271730731\\
3.42465753424658	-19.2619903573618\\
3.37837837837838	-18.9344011159713\\
3.33333333333334	-19.0448686175351\\
3.28947368421053	-18.6288913653253\\
3.24675324675324	-18.3498001206054\\
3.2051282051282	-18.1186183352529\\
3.16455696202532	-17.5172467003483\\
3.125	-17.7741505832244\\
3.08641975308642	-17.3686077375562\\
3.04878048780488	-17.1899044667483\\
3.01204819277108	-17.3807657477062\\
2.97619047619047	-17.0097204379793\\
2.94117647058823	-16.8506576926005\\
2.90697674418605	-16.779610412747\\
2.87356321839081	-16.6335054675237\\
2.84090909090909	-16.4101499282461\\
2.80898876404495	-16.4269002617672\\
2.77777777777778	-16.1837001029583\\
2.74725274725274	-15.8606458741839\\
2.71739130434783	-16.0001571827389\\
2.68817204301075	-15.7710023142638\\
2.6595744680851	-15.3049756333028\\
2.63157894736842	-15.205449977844\\
2.60416666666666	-16.0696254848842\\
2.57731958762886	-14.571277101432\\
2.55102040816327	-14.3745266582579\\
2.52525252525253	-15.2261960341983\\
2.5	-14.6840545171435\\
2.47524752475248	-14.3440882794842\\
2.42718446601942	-14.7650016500918\\
2.40384615384615	-13.8707877728474\\
2.38095238095238	-13.6134538391832\\
2.35849056603774	-13.6246937384793\\
2.33644859813084	-13.6903407674068\\
2.31481481481482	-13.6454073744761\\
2.29357798165137	-13.4606368595553\\
2.27272727272727	-13.156041925732\\
2.25225225225226	-13.7147204486027\\
2.23214285714285	-13.1706736052452\\
2.21238938053097	-12.9706612093509\\
2.19298245614035	-13.0898386111977\\
2.17391304347826	-12.9278759158894\\
2.1551724137931	-12.6874437134276\\
2.13675213675214	-12.7365428911753\\
2.11864406779661	-13.0055864921389\\
2.10084033613445	-12.4813864505106\\
2.08333333333334	-12.2583445162388\\
2.06611570247934	-12.2023645129785\\
2.04918032786885	-12.4749694214318\\
2.03252032520325	-12.3404942313888\\
2.01612903225806	-12.1292530303124\\
2	-12.008654439388\\
1.98412698412699	-12.171191689376\\
1.96850393700787	-11.9659752813898\\
1.953125	-11.6573089367983\\
1.93798449612403	-11.8085451084505\\
1.92307692307692	-11.7714560870817\\
1.90839694656488	-11.699052922495\\
1.89393939393939	-12.0584166974973\\
1.8796992481203	-11.7511473782964\\
1.86567164179105	-11.4776374104818\\
1.85185185185185	-11.4504915804892\\
1.83823529411764	-11.9755751824369\\
1.82481751824817	-11.3151671513812\\
1.81159420289855	-10.915447443632\\
1.79856115107913	-11.2482733111527\\
1.78571428571428	-10.459538420729\\
1.77304964539007	-11.0909583216124\\
1.76056338028169	-10.9453277466306\\
1.74825174825175	-10.9024418609262\\
1.73611111111111	-10.5349692930207\\
1.72413793103448	-11.005305254209\\
1.71232876712329	-10.7877190912392\\
1.70068027210884	-10.6341816400809\\
1.68918918918919	-10.6841583198643\\
1.67785234899329	-10.3160873708262\\
1.66666666666666	-10.4143151025155\\
};
\addlegendentry{2.4 GHz}

\end{axis}
\end{tikzpicture}%
\label{fig:s11_phase_sim1}}
\put(-202,85){\textcolor{black}{\small{\shortstack[l]{2.5~mm}}}}
\put(-202,22){\textcolor{black}{\small{\shortstack[l]{2.5~mm}}}}
 \caption{Insertion Loss optimal near 5:1 ratio}
\end{subfigure}
\\
\begin{subfigure}[t]{0.15\textwidth}
\centering
 {\includegraphics[width=\linewidth]{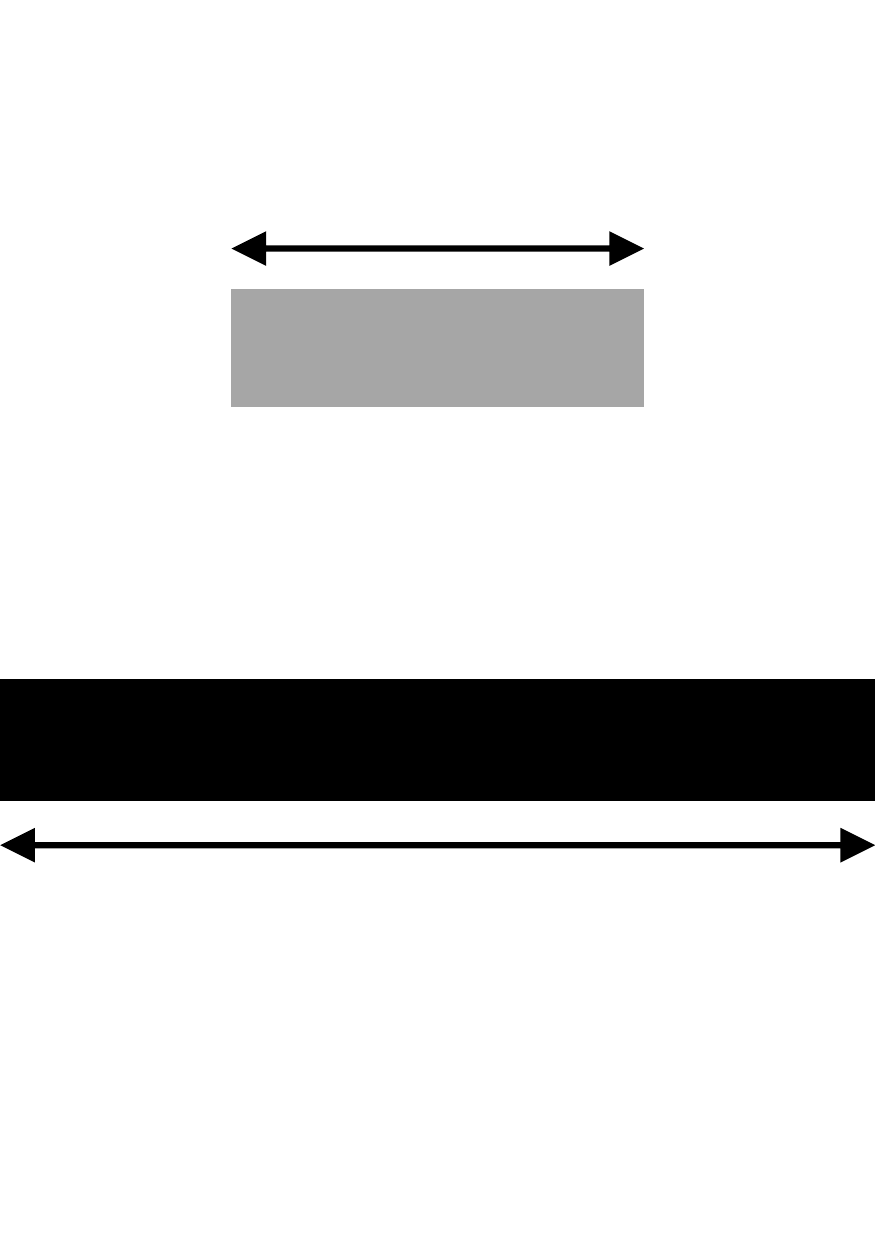}\label{fig:s11_mag_sim2}}
 \caption{Signal (gray) and Ground (black) traces}
\end{subfigure}
\begin{subfigure}[t]{0.35\textwidth}
\centering
 {
%
%
\definecolor{mycolor1}{rgb}{0.00000,0.44700,0.74100}%
\definecolor{mycolor2}{rgb}{0.85000,0.32500,0.09800}%
\begin{tikzpicture}

\begin{axis}[%
width=1.4in,
height=1in,
at={(1.297in,1.093in)},
scale only axis,
xmin=1,
xmax=9,
xlabel={Height to Width Ratio},
ymin=-60,
ymax=19,
ylabel style={at={(axis description cs:-0.25,0.425)},anchor=south},
ylabel={S11 Magnitude (dB)},
axis background/.style={fill=white},
xmajorgrids,
ymajorgrids,
legend style={at={(0.4,0.83)},anchor=west, nodes={scale=0.75, transform shape}, legend cell align=left, align=left, draw=white!15!black}
]
\addplot [color=blue, line width=1.5pt]
  table[row sep=crcr]{%
8.33333333333334	-11.5022452097954\\
8.06451612903226	-10.9263875580427\\
7.8125	-11.9760910244339\\
7.57575757575758	-12.3694555827187\\
7.35294117647059	-13.2475371500318\\
7.14285714285715	-13.131343221568\\
6.94444444444444	-14.2375290118683\\
6.75675675675676	-15.2541356092972\\
6.57894736842105	-16.6934897585912\\
6.41025641025641	-16.3820777803963\\
6.25	-17.4127583352144\\
6.09756097560976	-18.0161220729449\\
5.95238095238096	-18.995251336334\\
5.81395348837209	-20.9648790220529\\
5.68181818181818	-19.8720918090962\\
5.55555555555556	-23.5245132900616\\
5.43478260869566	-24.0757346868319\\
5.31914893617022	-26.6149856809849\\
5.20833333333334	-28.2097824508269\\
5.10204081632653	-30.6718041090696\\
5	-32.845839272523\\
4.90196078431373	-44.9828197083131\\
4.80769230769231	-34.1170045728364\\
4.71698113207547	-45.7257513884662\\
4.62962962962963	-44.7203068745458\\
4.54545454545455	-35.215217056572\\
4.46428571428572	-34.0325980368384\\
4.3859649122807	-30.4970838078741\\
4.31034482758621	-29.4504759759047\\
4.23728813559322	-24.6169174620513\\
4.16666666666666	-22.9940255129466\\
4.09836065573771	-23.7362461924666\\
4.03225806451613	-24.8300245045854\\
3.96825396825397	-24.1447515940631\\
3.90625	-20.6789186323696\\
3.84615384615385	-20.377130745304\\
3.78787878787879	-20.9581833090284\\
3.73134328358209	-21.2421466665012\\
3.6764705882353	-18.8571180143783\\
3.6231884057971	-18.0244898263027\\
3.57142857142857	-17.3579080135574\\
3.52112676056338	-17.4087044201381\\
3.47222222222222	-17.0573245880088\\
3.42465753424658	-16.3804318766027\\
3.37837837837838	-18.73941185605\\
3.33333333333334	-18.003381638368\\
3.28947368421053	-15.7378636746173\\
3.24675324675324	-15.6458118680424\\
3.2051282051282	-14.952394976907\\
3.16455696202532	-16.8384268567387\\
3.125	-15.3216953280307\\
3.08641975308642	-16.1509638637516\\
3.04878048780488	-14.7464764252282\\
3.01204819277108	-14.6829846673961\\
2.97619047619047	-13.9306274748111\\
2.94117647058823	-14.9450471717734\\
2.90697674418605	-14.4989027909163\\
2.87356321839081	-14.7457892229452\\
2.84090909090909	-13.1636545742879\\
2.80898876404495	-12.7993719161932\\
2.77777777777778	-12.8270055506871\\
2.74725274725274	-12.4039974650311\\
2.71739130434783	-12.4658128873427\\
2.68817204301075	-12.2078295786791\\
2.6595744680851	-11.8025790733612\\
2.63157894736842	-12.3485948164487\\
2.60416666666666	-12.0408867285353\\
2.57731958762886	-12.2225142256541\\
2.55102040816327	-12.7135310764389\\
2.52525252525253	-12.0064028140163\\
2.5	-12.185385527377\\
2.47524752475248	-11.8607297958951\\
2.45098039215686	-11.908132972604\\
2.42718446601942	-11.8624449006988\\
2.38095238095238	-10.9433918746554\\
2.35849056603774	-12.1968347483019\\
2.33644859813084	-11.2722984253638\\
2.31481481481482	-11.5125028785455\\
2.29357798165137	-10.3070785973491\\
2.27272727272727	-11.4652273970583\\
2.25225225225226	-10.8437948724397\\
2.23214285714285	-10.6483711030616\\
2.21238938053097	-10.0569676143213\\
2.19298245614035	-10.6029162145683\\
2.17391304347826	-10.4897063212233\\
2.1551724137931	-11.043111863313\\
2.13675213675214	-10.6778108363427\\
2.11864406779661	-10.1899337184216\\
2.10084033613445	-10.210310908682\\
2.08333333333334	-10.4722658435936\\
2.06611570247934	-9.99828542162819\\
2.04918032786885	-9.59189194951165\\
2.03252032520325	-9.99040351734213\\
2.01612903225806	-9.3433240677855\\
2	-9.46887890706011\\
1.98412698412699	-9.43377001617426\\
1.96850393700787	-9.51593748567501\\
1.953125	-9.23415667848765\\
1.93798449612403	-9.74010916157053\\
1.92307692307692	-9.3981025907717\\
1.90839694656488	-9.47829254277287\\
1.89393939393939	-9.3509629421219\\
1.8796992481203	-8.40849659905786\\
1.86567164179105	-8.68296585447114\\
1.85185185185185	-8.89522733423389\\
1.83823529411764	-9.97879854854208\\
1.82481751824817	-8.99107687860543\\
1.81159420289855	-8.68374182874723\\
1.79856115107913	-8.68667052281329\\
1.78571428571428	-8.51986584029598\\
1.77304964539007	-8.11317615800717\\
1.76056338028169	-9.09296833283447\\
1.74825174825175	-8.68090508240871\\
1.73611111111111	-8.60350891859925\\
1.72413793103448	-8.32769859388244\\
1.71232876712329	-7.78973698004679\\
1.70068027210884	-8.98660528803707\\
1.67785234899329	-7.63583843816431\\
1.66666666666666	-7.46985792914322\\
};
\addlegendentry{900 MHz}

\addplot [color=magenta, line width=1.5pt]
  table[row sep=crcr]{%
8.33333333333334	-16.9283393084955\\
8.06451612903226	-13.588998987488\\
7.8125	-13.8630446306587\\
7.57575757575758	-16.1788420533411\\
7.35294117647059	-16.4472850860093\\
7.14285714285715	-15.6801400786053\\
6.94444444444444	-15.4107837822597\\
6.75675675675676	-17.3780929054015\\
6.57894736842105	-19.5223251292105\\
6.41025641025641	-17.5904386793089\\
6.25	-16.4455985002628\\
6.09756097560976	-19.003825222704\\
5.95238095238096	-17.6587360024206\\
5.81395348837209	-22.8022974819167\\
5.68181818181818	-18.8219229360525\\
5.55555555555556	-26.3517962397325\\
5.43478260869566	-25.6719450163206\\
5.31914893617022	-23.9790225002735\\
5.20833333333334	-26.7778077104291\\
5.10204081632653	-27.1925151193842\\
5	-26.413964241983\\
4.90196078431373	-30.0442671250974\\
4.80769230769231	-28.8802971891236\\
4.71698113207547	-32.8701586700898\\
4.62962962962963	-30.3877970202703\\
4.54545454545455	-31.3015191073408\\
4.46428571428572	-26.6181911990975\\
4.3859649122807	-32.2646698437567\\
4.31034482758621	-40.4307254011617\\
4.23728813559322	-55.439369396973\\
4.16666666666666	-43.6567396045478\\
4.09836065573771	-38.947711020135\\
4.03225806451613	-36.2829607548979\\
3.96825396825397	-37.7861214784851\\
3.90625	-48.3584113397702\\
3.84615384615385	-31.471623782061\\
3.78787878787879	-38.3948633187923\\
3.73134328358209	-39.2110551670155\\
3.6764705882353	-30.007178187351\\
3.6231884057971	-27.8637508468374\\
3.57142857142857	-28.7026331127252\\
3.52112676056338	-27.6064975279381\\
3.47222222222222	-27.8388812127112\\
3.42465753424658	-26.5562029411941\\
3.37837837837838	-32.2801850315022\\
3.33333333333334	-33.0843054724365\\
3.28947368421053	-24.0751691553638\\
3.24675324675324	-24.2333734968304\\
3.2051282051282	-23.3709395696408\\
3.16455696202532	-37.5115907980255\\
3.125	-22.2512417673865\\
3.08641975308642	-30.294657447811\\
3.04878048780488	-22.0323091698037\\
3.01204819277108	-20.9693168766973\\
2.97619047619047	-21.0317079931899\\
2.94117647058823	-25.3498996294067\\
2.90697674418605	-24.204078035503\\
2.87356321839081	-27.1320897008335\\
2.84090909090909	-20.6376903256389\\
2.80898876404495	-19.9121677631742\\
2.77777777777778	-20.2646990809644\\
2.74725274725274	-20.2050140877962\\
2.71739130434783	-19.3556208109846\\
2.68817204301075	-19.2425868777715\\
2.6595744680851	-19.1462668469855\\
2.63157894736842	-19.2562784415618\\
2.60416666666666	-19.139215473173\\
2.57731958762886	-19.3222466293346\\
2.55102040816327	-34.3782714208813\\
2.52525252525253	-20.0505011214503\\
2.5	-22.0187405314221\\
2.47524752475248	-22.1004227242356\\
2.45098039215686	-17.9645964175366\\
2.42718446601942	-19.8167902623356\\
2.40384615384615	-17.6501955429227\\
2.38095238095238	-17.8993497231682\\
2.35849056603774	-21.913737912773\\
2.33644859813084	-17.2404393049173\\
2.31481481481482	-18.0712559578622\\
2.29357798165137	-17.6242237982704\\
2.27272727272727	-18.9826922473544\\
2.25225225225226	-16.7552175914014\\
2.23214285714285	-17.2794874051671\\
2.21238938053097	-16.4710842088709\\
2.19298245614035	-17.158411325648\\
2.17391304347826	-18.4474014042478\\
2.1551724137931	-17.1881430276536\\
2.13675213675214	-17.1348820507607\\
2.11864406779661	-16.1380903959668\\
2.10084033613445	-16.4432350834055\\
2.08333333333334	-17.4032765877968\\
2.06611570247934	-16.5165349500358\\
2.04918032786885	-15.5424689995892\\
2.03252032520325	-16.6513488927495\\
2.01612903225806	-15.1880719988959\\
2	-15.1434877125974\\
1.98412698412699	-15.4873683425634\\
1.96850393700787	-14.8305228167073\\
1.953125	-15.2691326155683\\
1.93798449612403	-16.4228683148054\\
1.92307692307692	-15.1837320558724\\
1.90839694656488	-15.8839776300263\\
1.89393939393939	-15.5834483511977\\
1.8796992481203	-14.2942850724099\\
1.86567164179105	-14.1121346666396\\
1.85185185185185	-15.256771262568\\
1.83823529411764	-15.8834549637333\\
1.82481751824817	-16.0039720951845\\
1.79856115107913	-14.0304425870193\\
1.78571428571428	-13.83428919431\\
1.77304964539007	-14.0879825160582\\
1.76056338028169	-15.3426077707464\\
1.74825174825175	-13.8649793596357\\
1.73611111111111	-13.7589550463405\\
1.72413793103448	-13.9495178858809\\
1.71232876712329	-13.5199675712731\\
1.70068027210884	-13.9662203802246\\
1.68918918918919	-13.7393539887013\\
1.67785234899329	-13.3254583965023\\
1.66666666666666	-13.5273827445912\\
};
\addlegendentry{2.4 GHz}

\end{axis}
\end{tikzpicture}%
\label{fig:s11_phase_sim2}}
\put(-202,85){\textcolor{black}{\small{\shortstack[l]{2.5~mm}}}}
\put(-200,22){\textcolor{black}{\small{\shortstack[l]{6~mm}}}}
 \caption{Insertion Loss optimal near 4:1 ratio}
\end{subfigure}
\\
\caption{HFSS simulation results: as the ground layer width is increased to allow for easier interfacing with the SMA connector, the ideal height:width ratio decreases from 5:1 to 4:1.}
\label{fig:hfss_sims}
\end{figure}
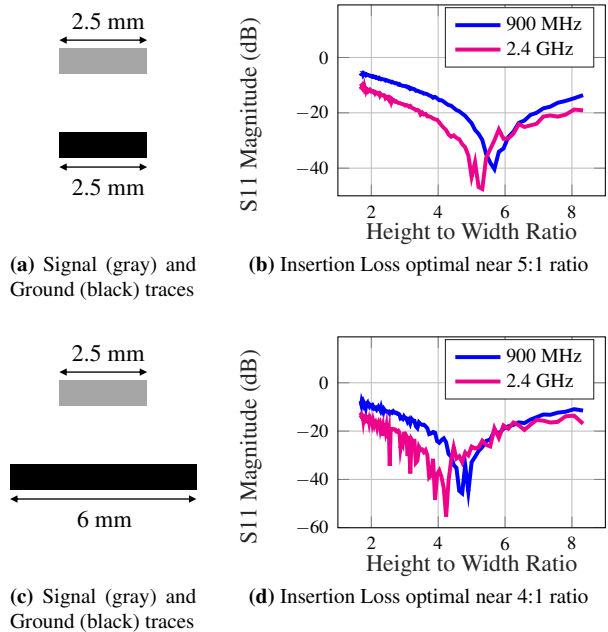

However, we notice some deviation from this ratio when the width of ground trace is increased to allow for easier interfacing with SMA connectors. We simulate the sensor in Ansys HFSS (Fig. \ref{fig:hfss_sims}) to determine this deviation, and observe that the ideal operating ratio shifts to about 4:1 instead of 5:1 when the width of ground trace is increased.


\subsection{Performance with distance}
\label{sec:dist_exp}
\noindent We also evaluate our sensor and wireless reader design over a range of distances. For this experiment, we place the TX antenna, sensor antenna and RX antenna along a straight line. The TX antenna is placed 4~m away from the RX antenna, and the sensor is moved from the mid-point, which is 2~m away from both to distances, closer to the RX antenna, and farther away from the TX antenna. The TX power is set to 10~dBm, and the center frequency for this experiment was 900~MHz. We can observe that the sensor gives accurate and satisfying phase stability of less than $1^{o}$ even at a distance of 1~m, 3~m from the RX/TX, and acceptable within $5^{o}$ performance even at the worst 2~m, 2~m distance from the TX/RX. These operating distances are comparable with previously shown evaluations with RFID based backscatter at 900~MHZ~\cite{pradhan2017rio}, which tested sensing at a maximum distance of 2~m from the RFID reader.

\begin{figure}[H]
\begin{subfigure}[t]{0.23\textwidth}
\centering
 {
%
%
\definecolor{mycolor1}{rgb}{0.00000,0.44700,0.74100}%
\begin{tikzpicture}

\begin{axis}[%
width=1.05in,
height=1.1in,
at={(1.297in,5.802in)},
scale only axis,
xmin=20,
xmax=200,
xlabel={Distance from RX (m)},
xtick={50,100,150,200},
xticklabels={0.5,1,1.5,2},
ymin=25,
ymax=40,
ylabel style={at={(axis description cs:-0.15,0.425)},anchor=south},
ylabel={SNR (dB)},
axis background/.style={fill=white},
xmajorgrids,
ymajorgrids,
legend style={legend cell align=left, align=left, draw=white!15!black}
]
\addplot [color=blue, mark=o, mark options={solid, mycolor1}]
 plot [error bars/.cd, y dir = both, y explicit]
 table[row sep=crcr, y error plus index=2, y error minus index=3]{%
20	36.3047714669224	0.951239554489743	0.951239554489743\\
30	33.0446428077265	1.24754437520351	1.24754437520351\\
40	31.052316363532	0.790784017284089	0.790784017284089\\
50	32.1379787015874	1.32924773222272	1.32924773222272\\
60	33.5111425977605	0.78900801286062	0.78900801286062\\
70	30.951516507736	1.07094409271818	1.07094409271818\\
80	30.8907149233683	2.24821933013813	2.24821933013813\\
90	29.632488655701	1.26759460423245	1.26759460423245\\
100	29.8466018927752	0.823299207767501	0.823299207767501\\
110	29.0213775137816	0.399764363720516	0.399764363720516\\
120	29.0573750385686	1.07359890681535	1.07359890681535\\
130	27.1505049448095	0.567889309248005	0.567889309248005\\
140	28.4641453645371	0.848902463530105	0.848902463530105\\
150	26.8500245763687	0.746239069590225	0.746239069590225\\
160	26.4305331197975	0.340311703421478	0.340311703421478\\
170	28.488439677289	0.614245742474132	0.614245742474132\\
180	27.4259901752832	0.801251622016995	0.801251622016995\\
190	29.1674384342918	0.952782814406291	0.952782814406291\\
200	26.8367512859819	0.793800356911312	0.793800356911312\\
};
\end{axis}

\end{tikzpicture}%
\label{fig:distance1}}
 \caption{Sensor SNR}
\end{subfigure}
\hfill
\begin{subfigure}[t]{0.23\textwidth}
\centering
 {
%
%
\definecolor{mycolor1}{rgb}{0.00000,0.44700,0.74100}%
\begin{tikzpicture}

\begin{axis}[%
width=1.05in,
height=1.1in,
at={(1.297in,1.093in)},
scale only axis,
xmin=20,
xmax=200,
xlabel={Distance from RX (m)},
xtick={50,100,150,200},
xticklabels={0.5,1,1.5,2},
ymin=0,
ymax=15,
ylabel style={at={(axis description cs:-0.15,0.425)},anchor=south},
ylabel={Phase Deviation (deg)},
axis background/.style={fill=white},
xmajorgrids,
ymajorgrids,
legend style={legend cell align=left, align=left, draw=white!15!black}
]
\addplot [color=blue, mark=o, mark options={solid, mycolor1}]
 plot [error bars/.cd, y dir = both, y explicit]
 table[row sep=crcr, y error plus index=2, y error minus index=3]{%
20	0.508040628051546	0.144270409891104	0.144270409891104\\
30	0.459724082497926	0.260469594552699	0.260469594552699\\
40	0.531928825061265	0.521953895801988	0.521953895801988\\
50	0.324222790051643	0.119712722558535	0.119712722558535\\
60	0.39813534490306	0.152326070112487	0.152326070112487\\
70	0.998053245890492	0.329995134490042	0.329995134490042\\
80	0.752207136017818	0.357162344783848	0.357162344783848\\
90	0.716621317360235	0.227432900976952	0.227432900976952\\
100	2.13474512579282	1.00484221323031	1.00484221323031\\
110	2.27906101583086	0.459066976550118	0.459066976550118\\
120	0.980719838062499	0.378111338840538	0.378111338840538\\
130	4.30247648344992	0.958721056801615	0.958721056801615\\
140	2.56878378740845	1.50521696230231	1.50521696230231\\
150	5.02859294530542	2.01073924440913	2.01073924440913\\
160	4.5527144050457	1.0469847308205	1.0469847308205\\
170	2.82922930944199	1.98406989204829	1.98406989204829\\
180	2.49717853918455	1.56308237809248	1.56308237809248\\
190	1.91249101105887	1.0191300184941	1.0191300184941\\
200	3.41833661007633	2.64015663308234	2.64015663308234\\
};

\addplot [color=black, line width = 2pt, dashed, forget plot]
  table[row sep=crcr]{%
20	5\\
200	5\\
};

\end{axis}
\end{tikzpicture}%
\label{fig:distance2}}
 \caption{Sensor Phase Std. Deviation}
\end{subfigure}
\caption{Testing \paperTitle over a range of distances}
\label{fig:distance}
\end{figure}
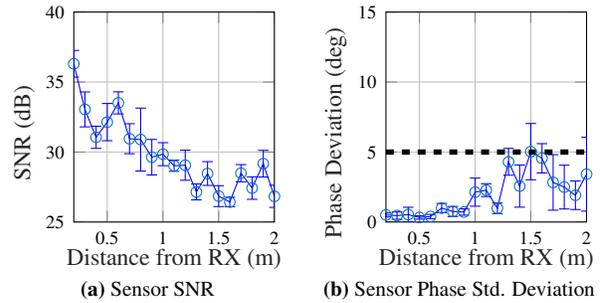

\vfill


\end{document}